\documentclass[11pt,a4paper]{article}
\usepackage[a4paper,margin=0.75in]{geometry}

\usepackage{setspace}
\usepackage{amsmath}
\usepackage[pdftex]{graphicx}
\usepackage{subfigure}
\usepackage{dcolumn}
\usepackage{epstopdf}
\usepackage{color}
\usepackage{upgreek}
\usepackage[hidelinks]{hyperref}
\usepackage{cite}
\usepackage{graphicx}

\makeatother

\date{}

\usepackage[nameinlink,noabbrev]{cleveref}
\crefname{figure}{Fig.}{Figs.}
\Crefname{figure}{Figure}{Figures}
\definecolor{JV}{RGB}{0, 0, 0}
\definecolor{myNewColor1}{RGB}{0, 0, 0}
\definecolor{myNewColor2}{RGB}{0, 0, 0}
\definecolor{MLColor}{RGB}{0, 0, 0}
\definecolor{JHColor}{RGB}{0, 0, 0}
\definecolor{OdRColor}{RGB}{110,10,117}
\definecolor{Red}{rgb}{1,0,0}

\DeclareRobustCommand{\jv}[1]{{\color{JV}#1}}

\DeclareRobustCommand{\ml}[1]{{\color{MLColor}#1}}
\DeclareRobustCommand{\jh}[1]{{\color{JHColor}#1}}

\makeatletter
\newcommand*{\addFileDependency}[1]{%
  \typeout{(#1)}%
  \@addtofilelist{#1}%
  \IfFileExists{#1}{}{\typeout{No file #1.}}%
}

\makeatother
\newcommand{\nref}[1]{Figure.~\ref{#1}}

\newcommand{\neqref}[1]{Eq.~\eqref{#1}}
\newcommand{\sref}[1]{Sec.~\ref{#1}}

\usepackage{titling}

\pretitle{\fontsize{15}{20}\selectfont\bfseries}
\posttitle{\par}

\begin{document}

\title{Designing corrugated surfaces to guide colloidal self-assembly}

\maketitle
\vspace{-2.7cm}

\begin{center}

Dinesh Kumar Sahu$^{1,2}$*, Jude Ann Vishnu$^{2}$, Lisa Shafroth$^{2}$, Martin Lenz$^{1,2}$, Olivia du Roure$^{2}$ and Julien Heuvingh$^{2}$*\\
$^{1}$Université Paris-Saclay, CNRS, LPTMS, 91405, Orsay, France\\
$^{2}$PMMH, CNRS, ESPCI Paris, PSL University, Sorbonne Université, Université Paris-Cité, 75005, Paris, France\\
$^{•}$*Emails : dinesh-kumar.sahu@espci.fr, julien.heuvingh@u-pariscite.fr
 
\end{center}

\begin{abstract}

The self-assembly of colloidal particles enables the creation of structured materials with programmable functionalities; however, controlling interaction specificity and aggregate morphology in a reversible and scalable manner remains a major challenge. Here, we investigate the selective depletion-induced self-assembly of 3D-printed flat polygonal colloids, where nanoscale surface topography is engineered through precise modeling in two-photon polymerization. By designing anisotropic lateral surfaces, we direct specific interactions that govern aggregate morphology, yielding dimers, chains, zigzag, and honeycomb structures depending on the surface configuration. The specificity of interaction is tuned by varying the length scale of the topographic surfaces, the depletant concentration and the ionic strength of the solution, revealing a transition from selective to non-selective aggregation regimes. The relative placement of lateral interacting surfaces on the colloids enables assembly into aggregates spanning a broad range of sizes, while tuning the interaction strength selectively stabilizes distinct structural motifs. We  demonstrate this interplay between geometric arrangement and interaction energy experimentally and corroborate through both theory and simulations for specifically hexagonal shaped colloids. This study establishes a versatile framework for programming colloidal interactions via micro-architectural design, offering new routes for fabricating reconfigurable and functional soft materials.

\end{abstract}

\section{Introduction}

At the microscale, Brownian motion, together with isotropic attractive interactions, makes self-assembly processes inherently stochastic and often leads to uncontrolled aggregation. To achieve programmable and reproducible assemblies, it is essential to engineer building blocks with interaction anisotropy that restricts bonding to specific, energetically favored configurations.\cite{whitesides2002self, manoharan2015colloidal, zhao2018assembly}. A wide range of experimental strategies have utilized this approach to produce ordered matter. In molecular systems, sequence programmable DNA hybridization drives highly specific assembly: DNA origami can be designed to fold and tessellate into arbitrary 2D and 3D architectures with nanometre precision, enabling the fabrication of addressable lattices and functional nanodevices \cite{rothemund2006folding, torring2011dna}. At larger length scales, colloidal systems provide an excellent platform for studying how shape, surface patterning and interaction anisotropy determine assembly pathways. For example, patchy colloids with chemically distinct or DNA-functionalized patches behave as colloidal analogues of atoms with programmable valence and directional bonds, leading to open lattices and complex crystals when patches are placed with controlled geometry and binding specificity \cite{wang2012colloids, glotzer2007anisotropy, shelke2025self}. Shape complementarity has also been exploited in lock-and-key colloids: particles with complementary cavities and protrusions selectively bind, producing reversible, orientation-specific assemblies reminiscent of molecular recognition \cite{sacanna2010lock, sacanna2011lock}.

Among the various interaction mechanisms, depletion interactions have proven exceptionally versatile due to their tunability and reversibility in colloidal systems \cite{asakura1958interaction, vrij1976polymers, wang2015colloidal, lekkerkerker2011stability}. The concept, originally formulated by Asakura and Oosawa, describes an effective attraction between large colloids induced by the osmotic pressure imbalance of smaller, non-adsorbing depletants that are excluded from a thin shell surrounding each colloid \cite{asakura1958interaction}. When two colloids approach within a distance smaller than twice the depletant radius, the overlap of exclusion zones increases the available free volume for the polymers, producing an entropic driving force that pushes the colloids together. The effective depletion potential between two colloids, immersed in a depletant solution of concentration $c$, is given by
\begin{equation}
U_{\mathrm{dep}} \approx - c k_{\mathrm{B}}T \Delta V  \tag{1} 
\end{equation}
where $\Delta V$ is the overlap volume of the excluded regions and $k_{\mathrm{B}}T$ is the thermal energy scale \cite{asakura1958interaction, vrij1976polymers}. This linear relationship between interaction strength and polymer concentration has provided a foundation to understand entropic assembly across diverse colloidal systems. However, more recent developments have extended the Asakura and Oosawa model to account for anisotropic particle shapes, polymer–polymer interactions, and surface roughness effects, which introduce directionality into the otherwise isotropic depletion potential \cite{lekkerkerker2011stability}.

A significant development toward physically encoded interaction specificity is the realization of selective depletion interactions, in which local surface morphology and particle geometry modulate the magnitude and spatial distribution of depletion-induced attractions \cite{sacanna2013engineering, petukhov2017entropic, zhang2023depletion}. The local curvature of a colloid surface mediate the strength of depletion interactions with the opposing colloid surface, as geometrically complementary surfaces will maximize the excluded volume loss when the two colloids interacts. Colloids possessing flat surfaces will preferentially interact through these surfaces as is the case for the historically relevant aggregation of platelets \cite{lekkerkerker2011stability} and artificial colloids \cite{rossi2011cubic, young2012assembly, tigges2016hierarchical, mayarani2025lifetime}. The presence of a locally convex and a locally concave surface will likewise create directional preferential depletion interactions if the two surfaces are complementary. This as been used to create patchy particles interacting like atoms \cite{sacanna2010lock}, or long range assembly \cite{wan2025curvature}.
Early work by Zhao \cite{zhao2007directing, zhao2008suppressing} and Badaire \cite{badaire2008experimental}
demonstrated that locally roughened surfaces can suppress depletion attraction if the asperities length scale is larger than the depletant size. This property has been used to
effectively encode “patchiness” into colloidal particles \cite{sacanna2012magnetic, kraft2012surface, kamp2016selective, wolters2015self}. Flat–Flat contacts yield strong depletion attraction, whereas rough–rough contacts are entropically unfavorable, resulting in interaction selectivity analogous to directional chemical bonding. 
In these studies, surface roughness was generally introduced in a qualitative fashion, without precise control over the detailed geometry, periodicity, or amplitude of the surface features. As a result, the ability to systematically tune interaction specificity and strength through controlled surface architecture has remained largely unexplored. In particular, ideally corrugated surfaces with periodic asperities could in principle lead to an increase in excluded volume overlap by allowing complementary surfaces to interlock \cite{zhao2007directing}. Developing approaches that enable precise and tunable control over directional interactions are therefore crucial for transforming colloidal self assembly into a predictive and programmable design strategy capable of producing finite, structurally defined, and hierarchically organized materials.

Recent advances in micro- and nanofabrication, especially two-photon polymerization technique enable fabrication with submicron control \cite{carlotti2019functional}. Recent studies have emphasized its capability to fabricate complex microstructures and to tune surface morphology \cite{greiner2012micro, serien2015fabrication}. However, the extent to which fabrication parameters govern sub-micrometer surface morphology, and how such controlled roughness directly translates into interaction anisotropy and specificity, has not yet been systematically investigated. In this study, we utilize two-photon polymerization based engineering to systematically design and control selective depletion interactions in polygonal shaped colloids. By tuning the design parameters during fabrication, we generate colloids with controlled combinations of flat (favourable) and corrugated (unfavourable) sides. Using depletion-induced attractions between these colloids, we demonstrate tunable assembly from directional dimers and one-dimensional chains to two-dimensional honeycomb lattices. Moreover, by modulating surface patterning and interaction strength, we identify the regimes where specific interactions dominate over non-specific aggregation. These findings establish a design framework for directing self-assembly through surface topography and colloidal thermodynamics, bridging top-down fabrication precision with bottom-up emergent organization.

\section{Results}

\subsection{Selective interaction between colloids}

\begin{figure}[h]
\centering
\includegraphics[scale=0.925]{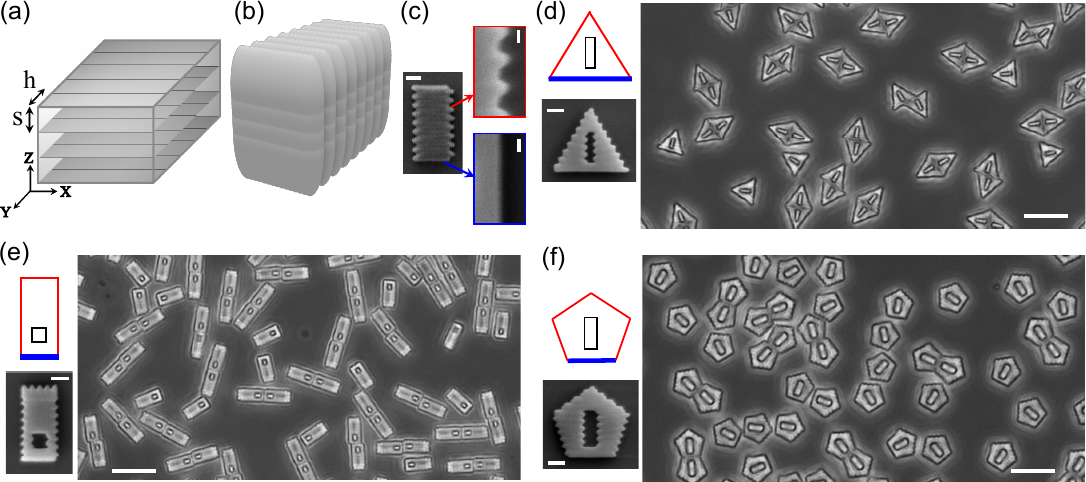}
\caption{Selective depletion interactions in 3D-printed polygonal colloids. (a) Schematic illustration depicting the layering and patterning of a rectangular prism shaped particle, highlighting the slicing distance ($s$) and hatching distance ($h$). (b) Visualization of the printed particle showing the discrete ellipsoidal voxels, the fundamental subunits that determine the spatial resolution and surface topography of the printed structure. (c) scanning electron microscopy (SEM) image of a rectangular shaped colloid printed along the hatch lines parallel to width (Scale bar = 1 $\mu m$). The highlighted images show clear distinction between flat and corrugated sides (Scale bar = 300 $nm$). (d–f) Design schematics and corresponding SEM images of triangular, rectangular, and pentagonal colloids. The blue colour denote sides engineered flat to promote favourable depletion attraction, whereas the red colour represent corrugated, non-interacting sides. The hole serves as a fiducial marker to identify the direction of the interacting sides (blue). The accompanying optical micrographs demonstrate dimeric self-assembly, confirming selective attraction between designed interfaces. Scale bars on SEM and microscopy images are 1 $\mu m$ and 10 $\mu m$, respectively.}
\label{fig:Fig.1}
\end{figure}

In the context of 3D printing, 
three-dimensional objects are sliced into a series of equidistant planes, each composed of equidistant lines. The separations between such consecutive planes and lines are defined as the slicing distance $s$ and hatching distance $h$, respectively (\nref{fig:Fig.1}(a)). In two-photon polymerization, a tightly focused pulsed laser beam is scanned along those hatching lines to fabricate 3D structures by triggering local polymerization of the photoresist (IP-L 780, Nanoscribe GmbH). The ellipsoidal focal volume of the laser, commonly termed as voxel, serves as the fundamental building block of the printed structure and plays a crucial role in defining its spatial resolution and surface topography. A schematic representation of 3D printing of \nref{fig:Fig.1}(a) is shown in \nref{fig:Fig.1}(b). Printing a polygonal particle with this method thus creates faces along the direction of printing that are expected to be flat, while faces perpendicular or oblique to the direction of printing might keep the geometry of the last voxels printed along the lines. We printed polygonal particles with a hatching distance of 300nm, and observed with scanning electron microscopy (SEM) that faces along the direction of printing are flat, whereas the others are corrugated (\nref{fig:Fig.1}(c)). Moreover, the length scale of the corrugation is found to be nearly double the hatching distance due to the alternative scanning between consecutive hatching lines (\sref{sec:dir_hatching}). We employed this approach to modulate the interactions between the colloids.

\vspace{0.2cm}

\begin{figure}[h]
\centering
\includegraphics[scale=0.88]{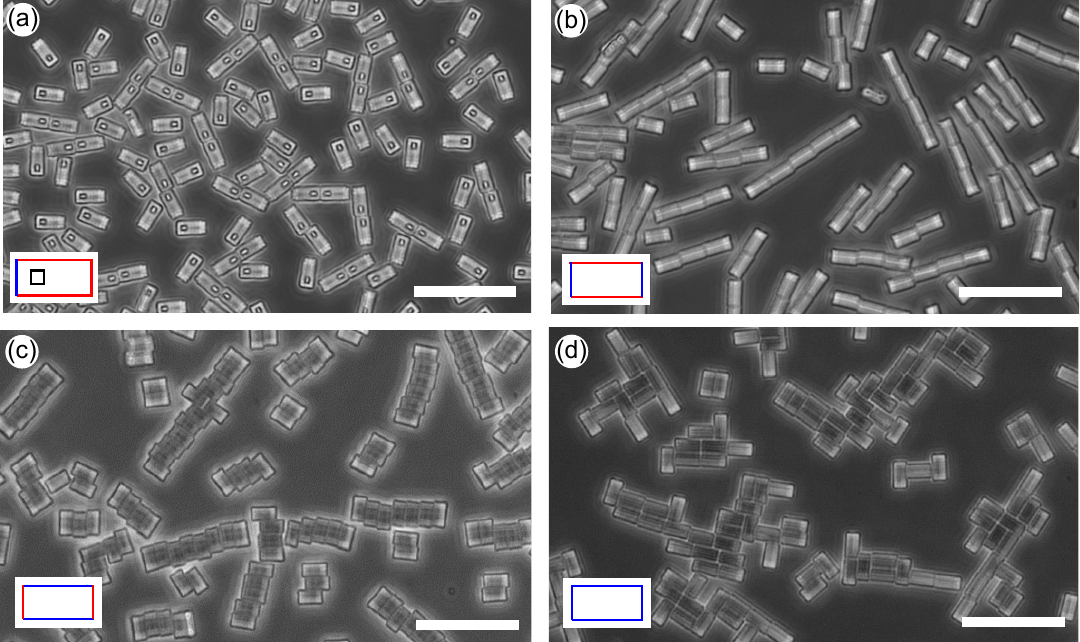}
\caption{Controlled self-assembly of rectangular colloids mediated by depletion interactions on selected sides. (a) Dimer formation driven by selective attraction on a single short side of the colloid, yielding directional pairwise association. (b) One-dimensional elongated aggregates emerging from specific depletion interactions localized on the shorter sides, leading to anisotropic chain growth (Movie-S1). (c) Hierarchical stacking of preformed linear aggregates through interactions along the longer sides, resulting in extended fibrous architectures. (d) Non-specific clustering arising when all sides are rendered flat, producing amorphous multi-colloidal aggregates. Insets schematically illustrate the designed surface selectivity, where blue sides represent favourable (attractive) surfaces and red sides denote unfavourable (non-interacting) faces. Scale bar is 20 $\mu m$.}
\label{fig:Fig.2}
\end{figure}

We first examine the self-assembly of flat polygon microstructures, including triangular, rectangular, and pentagonal shaped colloids, fabricated using distinct hatching patterns (\sref{sec:fabrication} and \sref{sec:hatching_strategy}) that produces a single flat lateral surface per colloid. In this approach, the hatch lines are oriented parallel to one selected side of the polygon, indicated by the blue face in the schematics (\nref{fig:Fig.1}(d–f)). To facilitate the identification of the flat face and orientation of colloids during the assembly process, a rectangular hole was incorporated inside each colloid. Corresponding SEM images confirm the distinct surface morphologies, clearly resolving the flat and corrugated lateral faces. To enable the release of fabricated particles from the substrate, a sacrificial layer (2~wt$\%$ of Polyacrylic acid, PAA) is deposited prior to colloid fabrication. We introduce depletion effects by choosing polyethylene glycol (PEG, 0.6~MDa) at a concentration of 10~$\mu$g/ml as depletant of size $R_h =$ 52 $nm$, together with a nonionic surfactant (Tergitol-NP10) and NaCl (20~mM), to regulate inter-colloid interactions. Upon gentle drop-casting of the depletant solution onto the printed arrays, the colloids detach from the substrate as the underlying PAA layer dissolves. They subsequently undergo Brownian diffusion \jh{while staying close to the substrate}. We measures the translational diffusion coefficient of 0.033 $\pm0.002~\mu m^{2}s^{-1}$ for rectangular colloids (\sref{sec:diffusion}). \jh{Subject to} diffusion, colloids encounter one another and interact preferentially through their flat lateral faces. As a result, the system predominantly forms dimers. The relative orientation of colloids within each dimer, identified by the position of the central hole, confirms the bonding selectivity along the flat faces. The inter-colloid bonds remain stable under the current conditions while the colloids exhibit relative sliding motion along the contact, a characteristic signature of bond fluctuations \cite{mayarani2025lifetime, zhao2007directing}. This selective interaction results in the formation of well-defined aggregates governed by surface-encoded interaction specificity. Interactions through the corrugated faces are strongly suppressed, despite their characteristic length scale exceeding the effective depletant size. In principle, one might anticipate interlocking between two \jh{indentical corrugated faces, which would increase their interaction energy. However, we observed a dramatic reduction of the attraction between corrugated surface.} The origin of this reduction is discussed in detail in the Discussion section.

\subsection{Hatching pattern governs directional assembly}

\jh{ To explore the relationship between surface topography and aggregate morphology, we employed various hatching patterns to observe} their self-assembled aggregates (\sref{sec:hatching_strategy}). For sake of simplicity, we implemented this approach by using rectangular colloids, wherein the hatching angle was systematically varied. When only one lateral side of the colloid was made flat and the remaining faces were corrugated, the colloids preferentially assembled into dimers (\nref{fig:Fig.2}(a), consistent with \nref{fig:Fig.1}(e)). When the two shorter sides were rendered flat by orienting hatch lines parallel to the colloid width, the colloids promoted side-to-side attachment, leading to the formation of one-dimensional, chain-like assemblies (\nref{fig:Fig.2}(b)). Conversely, making the longer side flat by orienting the hatching lines parallel to the particle length reversed this behaviour, resulting in end-to-end attachment through the elongated flat sides (\nref{fig:Fig.2}(c)). When the hatching lines are kept parallel to all the sides such as contour lines, making all lateral surfaces flat, the colloids exhibit non-specific, isotropic aggregation without directional preference (\nref{fig:Fig.2}(d)). These observations collectively demonstrate that the hatching strategy \jh{can }serve as a critical design parameter for tailoring specific, directional interactions among micro-fabricated colloids.

\subsection{Conditions for specific interaction}

\begin{figure}[h]
\centering
\includegraphics[scale=0.88]{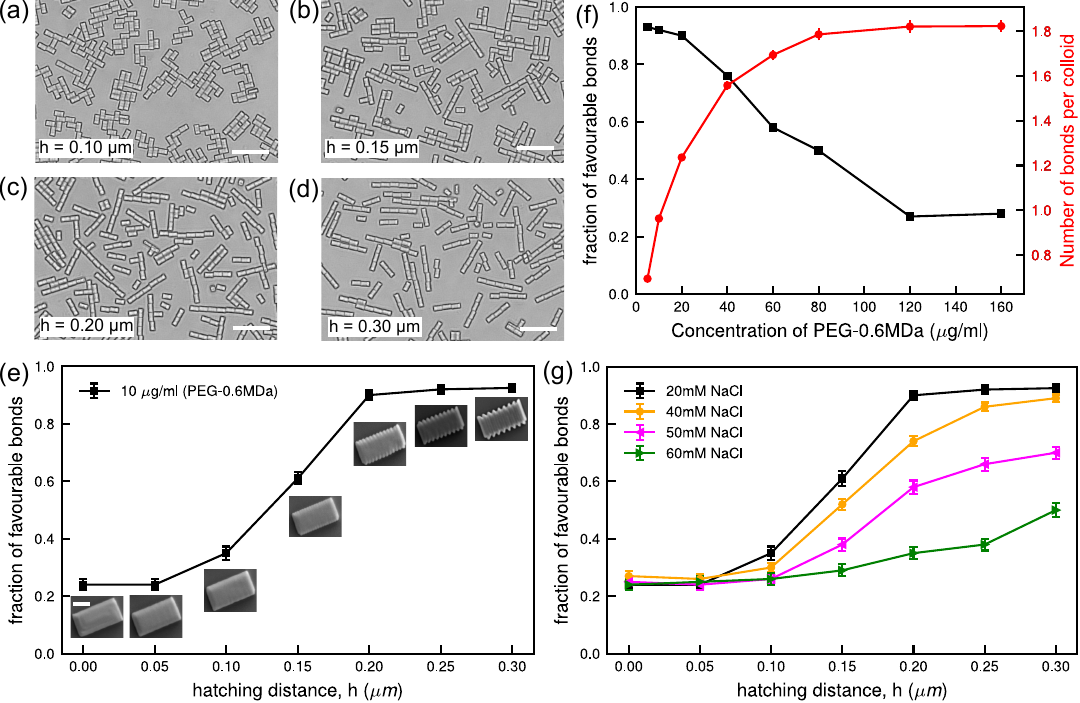}
\caption{Range of parameters enabling specific depletion-mediated interactions in rectangular colloids. (a-d) Representative optical micrographs showing distinct self-assembled morphologies of colloids fabricated at different hatching distances ($h$), highlighting the influence of sub-microscale surface topography on selective binding. Scale bar is 20 $\mu m$. (e) Quantitative variation of interaction specificity as a function of the hatching distance at 20mM salt concentration and 10 $\mu$g/ml of PEG-0.6 MDa. Corresponding SEM images of printed colloids (insets) demonstrate differences in surface morphology (Scale bar = 1 $\mu m$). (f) Dependence of interaction specificity on the overall interaction energy, modulated by the depletant concentration. The black symbols denote the measured specificity parameter, while the red curve represents the average fraction of contacts among the colloids. The observed decline in specificity with increasing interaction strength suggests the onset of non-selective, energetically favourable associations. (h) Variation in the fraction of specific interactions with increasing salt concentration at constant depletant concentration (10 $\mu$g/ml of PEG-0.6MDa), demonstrating that enhanced electrostatic screening diminishes directional selectivity and promotes non-specific aggregation.}
\label{fig:Fig.3}
\end{figure}

The specificity of depletion-induced interactions among the colloids is expected to depend not only on the presence of favourable and unfavourable contact regions but also on their interaction strengths, which can be influenced by both the colloid design parameters and the \jh{chemical composition of the suspending solution}. To systematically explore the range of conditions under which interaction specificity is preserved, we again employed rectangular-shaped colloids as a model system. Colloids were fabricated with a range of hatching distances ($h$) applied parallel to their shorter sides, and the resulting self-assembled structures were analyzed to assess directional aggregation behaviour. Based on the orientation of the hatching pattern, bond formation along the shorter sides are classified as favourable, while bonds formed along the remaining sides are considered unfavourable. We quantify the interaction specificity as the fraction of favourable bonds relative to the total number of bonds formed. Representative micrographs of the aggregates (\nref{fig:Fig.3}(a-d)) reveal a clear transition from isotropic to directionally specific assembly as the hatching distance $h$ increases. A systematic variation of $h$ from 0 to 0.30 $\mu m$ in increments of 0.05 $\mu m$ is shown in \nref{fig:Fig.3}(e) (black squares), obtained for a depletant concentration of 10 $\mu$g/ml (PEG-0.6 MDa) at 20 mM of NaCl. At small $h$ values, colloids aggregate without directional preference, forming isotropic assemblies (\nref{fig:Fig.3}(a) and (b)). As $h$ increases, the degree of interaction specificity increases, reaching a maximum value of approximately 0.9 at $h = 0.20$~$\mu m$, accompanied by pronounced directional assembly (\nref{fig:Fig.3}(c-d)). 

\jh{At the smallest hatching distance of 0.05~$\mu m$, the corrugation wavelength is presumably smaller than the depletant diameter (104 nm). In this situation, we expect the surface topography to be irrelevant for depletion attraction leading to lower specificity in the interaction at smaller hatching distances (\nref{fig:Fig.3}(e)). The corrugation wavelength increases with hatching distance, resulting in a consistent reduction in the depletion interaction along the corrugated surface. The maximum specificity of the corrugated surface is however achieved for hatching distances much larger than the depletant size (0.2~$\mu m$, corresponding to a corrugation wavelength of 400 nm). 
We provide in the discussion an explanation for this observation based on small variations of the surface geometry (\sref{sec:Discussion}). The scenario changes for colloids fabricated in unidirectional laser scanning mode where the corrugation wavelength equals the hatching distance, resulting in maximum interaction specificity at approximately double the hatching distance required in bidirectional scanning (\sref{sec:dir_hatching}).}

Next, we investigated the influence of interaction strength on the specificity of depletion-driven assembly by varying the depletant concentration of PEG-0.6~MDa at 20~mM of NaCl for colloids printed with $h$ = 0.30~$\mu m$. \jh{At higher depletant concentration (40 to 160 $\mu$ g/ml), we observed} a progressive reduction in interaction specificity. The interaction specificity reaches a minimum value of approximately 0.23 at concentrations of 120 $\mu$g/mL. (\nref{fig:Fig.3}(f)) \jh{and above. We quantified the average fraction of total bonds per colloid, including both favourable and unfavourable contacts (red curve in \nref{fig:Fig.3}(f))}. This quantity increases monotonically with depletant concentration and approaches saturation at higher concentrations. \jh{ We interpret the reduced bond fraction at lower concentrations as an effect of weaker depletion attraction, which reduces the number of colloids forming aggregates.} The observed reduction in specificity at elevated concentrations \jh{most likely arises from the increase of the interaction between corrugated surfaces.} These results indicate that, at higher interaction energies, previously unfavourable contacts become energetically favourable, thereby diminishing the overall directional specificity of the assembly.

\jh{As an} alternative mean of modulating the interaction strength \jh{between the colloid surfaces}, we changed the salt concentration in our solution. Increasing NaCl concentration reduces the Debye length but increases the double layer repulsion between charged surfaces \cite{israelachvili2011intermolecular}.   
To examine this effect, we fabricated rectangular shaped colloids with different hatching distances at let them assemble at increasing salt concentration and  constant depletant concentration (10 $\mu$g/ml of PEG-0.6MDa). As shown in \nref{fig:Fig.3}(g), increasing the salt concentration leads to a systematic reduction in fraction of favourable interactions. This observation is consistent with \jh{an increase of the interaction between corrugated surfaces, which promotes unfavourable interactions among the colloids.} 

\subsection{Designing complex aggregates}

\begin{figure*}[h]
\centering
\includegraphics[scale=0.9]{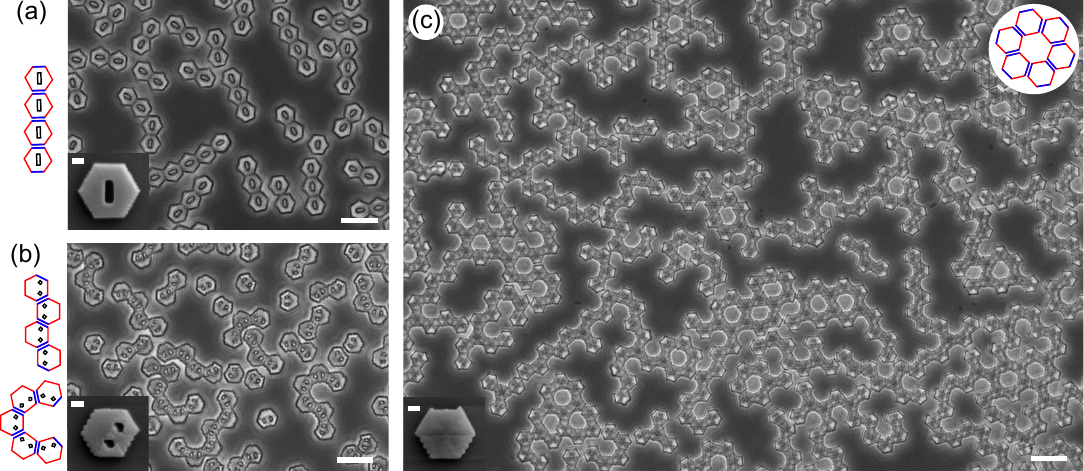}
\caption{Self-assembly of hexagonal colloids with selective interactions. (a) Hexagonal colloids engineered with two opposite flat (favourable) sides self-assemble into linear chain-like aggregates, driven by selective depletion attractions along the flat sides. (b) When two flat sides are separated by one corrugated (unfavourable) side, the particles organize into zigzag or circular configurations, reflecting the geometric constraints imposed by asymmetric interaction sides. (c) Introducing alternating flat and corrugated sides enables the formation of honeycomb-like lattices, demonstrating two-dimensional organization through selective depletion interactions. The corresponding SEM micrographs (insets) reveal the flat and corrugated sides of each designed hexagonal colloids. Schematic illustrations of the designed particle geometry and resulting aggregates are shown alongside, where blue denotes favourable and red denotes unfavourable interaction sides. Scale bars on SEM and optical microscopy images are 1 $\mu m$ and 10 $\mu m$, respectively.}
\label{fig:Fig.4}
\end{figure*}

Having established control over one-dimensional assembly through selective surface interactions, we next aimed to realize complex two-dimensional architectures with multiple distinct morphologies. This requires a geometry offering several independently addressable sides, enabling different combinations of attractive and non-attractive interfaces to encode specific interactions. To achieve this, we employed hexagon-shaped colloids, whose six-fold symmetry provides a versatile platform for generating diverse assemblies including chains, zigzag structures, rings, and honeycomb networks simply by altering which facets are functionalized. To realize such selective interactions, we adopted a hybrid surface-patterning strategy that integrates both hatching and contour-based fabrication techniques within the design of individual colloids (\sref{sec:hatching_strategy}). Contour-base hatching produces flat sides by aligning the hatch lines parallel to every sides. Each individual hexagonal colloid was printed in two sequential steps. In the first step, lateral sides designated to be unfavourable were patterned with perpendicular or inclined hatching lines at a spacing of $h = 0.30,\mu m$, to generate a corrugated surface that suppresses depletion-induced attraction. In the second step, the remaining faces, those designated as favourable were patterned using contour hatch lines, resulting in flat sides that promote specific interactions. 
The result was a single hexagonal colloid featuring a designed arrangement of flat and rough sides. We first studied the simplest scenario, consistent with our earlier findings in \nref{fig:Fig.1}, where particles possessing only one \jh{flat and five corrugated} sides preferentially form dimers. Following the same design principle, we fabricated hexagonal colloids with two opposing flat sides and the remaining four sides corrugated. These colloids assembled end-to-end along their flat sides, giving rise to linear chain-like aggregates (\nref{fig:Fig.4}(a)), analogous to the fibrous structures observed in rectangular colloids.

When the positions of the two flat sides were modified such that the \jh{flat} sides were no longer opposite but separated by one or more \jh{corrugated} sides, the assembly pathway changed dramatically. In the case where two favourable sides are separated by one unfavourable side, the colloids self-assemble into zigzag or occasionally curved aggregates, demonstrating that the interactions remain strictly confined to the flat sides (\nref{fig:Fig.4}(b)). This outcome confirms the high degree of directional specificity introduced through surface patterning. Finally, when the surface design incorporated alternating flat and \jh{corrugated} sides around the hexagon, the colloids \jh{assembled} into honeycomb-like 2D networks (\nref{fig:Fig.4}(c)). The resulting structures emerged through progressive colloidal aggregation, in which multiple nuclei of small honeycomb domains formed independently across the substrate and subsequently expanded through successive inter-cluster aggregation. The sustained presence of numerous spatially distributed honeycomb domains indicates that, as aggregates increase in size, their mobility decreases. Typical mismatch in honeycomb domain shapes also contributes in limiting further large-scale coalescence and stabilizing distinct mesoscale domains within the system. These results collectively highlight the generality and tunability of the surface-engineering strategy, demonstrating that precise spatial control over surface roughness and interaction anisotropy enables the programmable assembly of micro-structures into complex 2D morphologies from linear chains to zigzag and honeycomb networks through depletion-mediated interactions.

\subsection{Controlling self-limiting aggregation via interaction energy} \label{sec:Sec2_6}

\begin{figure*}[h]
\centering
\includegraphics[scale=0.925]{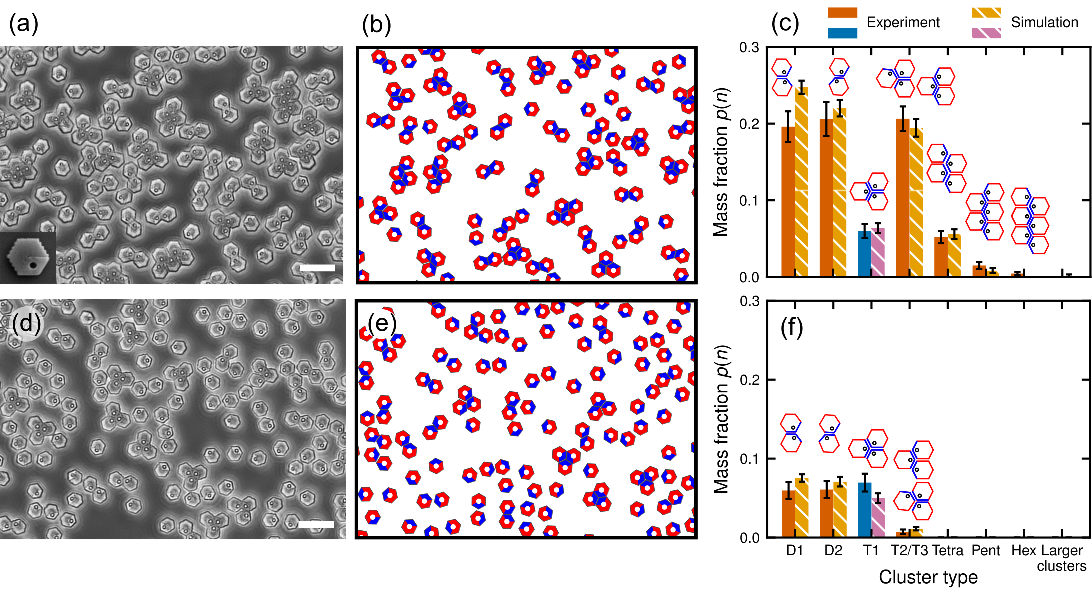} 
\caption{Tuning interaction energy to control the self-assembly of hexagonal colloids. (a) Under the condition of specificity (10 $\mu$g/ml of PEG-0.6MDa and 20mM NaCl), hexagonal colloids possessing two consecutive favourable sides self-assemble into extended aggregates in contrast to the expected self-limiting trimers T1. Inset shows the SEM image of a colloid. (d) Controlled aggregation with higher probability of self-limited trimers at low salt concentration (15mM). Scale bar is 10 $\mu m$. (b) and (e) Snapshots of aggregates from the simulation at the experimental time of 4 hours. (c) and (f) Statistics of aggregate size corresponding to (a,b,d,e) illustrating a change in a broad population of higher-order clusters to a narrow population dominated by self-limited trimers in both experiment and simulation.}
\label{fig:Fig.5}
\end{figure*}

\jh{
Hexagonal colloids bearing two consecutive flat sides are of particular interest because they depart from the specific binding behavior described above. Such colloids are expected to form self-limiting trimeric aggregates, in which each colloid engages both flat sides in favourable contacts, maximizing the number of favourable bonds and producing an energetically preferred configuration in which no attractive sides remain available for further growth.} Experimental observations however revealed \jh{a frequent assembly of the colloids into larger (tetrameric, pentameric,..) aggregates. This deviation can be understood} by considering the formation routes of dimers and trimers (\nref{fig:Fig.5}(c)). \jh{ Initially, dimers can form with their free favourable sides either on the same (D1) or opposite (D2) side of the dimer. 
A third colloid approaching the D2 dimer cannot close the trimer to form a self-limiting geometry, whereas the D1 can if the orientation of its flat sides align with the free flat sides of the dimer.} Consequently, three distinct types of trimeric aggregates emerge: (T1) a self-limiting trimer, where all favourable sides are mutually in contact, (T2) trimers, where one favourable side remain available for further attachment and (T3) trimers with two favourable sides available for interaction. The latter two configurations possess residual open binding sides that allow further growth into tetramers, pentamers, hexamers, and higher-order aggregates (\nref{fig:Fig.5}(a)). This is confirmed by the mass-fraction of aggregate sizes $p(n)$ (\nref{fig:Fig.5}(c)), which is defined to be the number of colloids in a given cluster of size $n$, divided by the total number of colloids in the system. Interestingly, the experimentally observed \jh{fraction of self-limited trimers to the total number of colloids engaged in trimeric or larger aggregate is 0.17, close to the value of 1/6 calculated by the assembly route from monomers to larger aggregates detailed above.} 

To obtain the self-limiting trimer aggregate, we aimed at suppressing the formation \jh{of the extended intermediates, by lowering the interaction energy between colloids. The goal is to make aggregations between two sides transient, while the T1 trimer would be stabilized by the cooperative binding of the three pairs of flat sides it exhibits}. We thus reduced the salt concentration to 15~mM (at constant depletant concentration), thereby enhancing the \jh{double layer} electrostatic repulsion between colloids.
\jh{We indeed observed in this condition that dimers and trimers in T2 and T3 configurations only appeared transiently, before separating over the course of minutes (Movie-S2).} 
The resulting aggregate population \jh{at 4 hours} showed a noticeable increase in the proportion of self-limiting trimers \jh{compared to the total number of aggregates}, accompanied by a complete disappearance of \jh{aggregates larger than trimers} (\nref{fig:Fig.5}(d) and (f)). Notably, under these weaker interaction conditions, monomers were more prevalent in the system, \jh{confirming}
that \jh{a single interaction between flat sides} was insufficient to reliably stabilize dimer formation. Nonetheless, the selective stability of  \jh{well}-oriented trimers throughout the experiment highlights the ability to program self-limiting assembly pathways through precise tuning of inter-colloid potentials.

Although reducing the salt concentration leads to a higher yield of self-limited trimers relative to dimers and \jh{larger} 
clusters, a comparison of the statistics at 4 hours for both the 20~mM salt and 15~mM salt cases (\nref{fig:Fig.5}(c) and (f)) reveals that the mass fraction of \jh{self-limited} trimers (T1) remains nearly the same in the two cases. 
Nevertheless, we expect the lower-salt system to favour a higher yield of 
T1 trimers as equilibrium is approached, because the weaker effective attraction should make bonds more reversible and thereby facilitate the rearrangement of kinetically trapped T2 and T3 aggregates toward the thermodynamically preferred T1 structure. Experimental limitations precluded us from observing the assembly evolution over the timeframe necessary to reach equilibrium. Hence, to examine this, we performed molecular dynamics simulations where the hexagonal colloids are modeled as rigid-bodies composed of discrete beads (\sref{sec:poly_model} and \sref{sec:hex_colloids}). In our model the two consecutive flat sides are assigned attractive interactions and the rough sides interacts purely via hard-core repulsion (indicated by red sides in \nref{fig:Fig.5}(b) and (e)). We performed a series of simulations over a range of attractive interaction strengths, and the resulting aggregate morphologies at the end of 4 hours are shown in \nref{fig:Fig.S8}(a)-(f).

\begin{figure*}[h]
\centering
\includegraphics[scale=0.925]{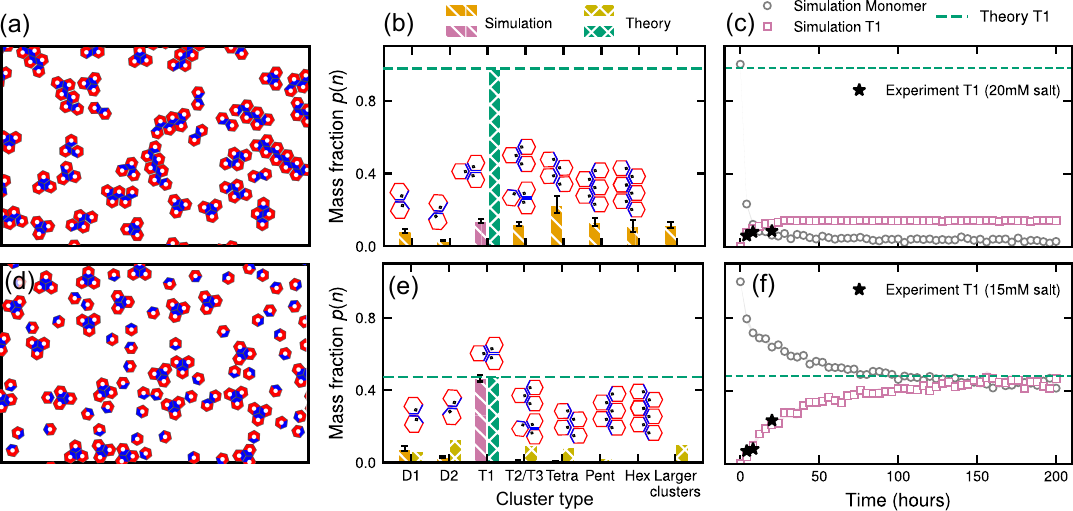} 
\caption{Comparison of equilibrium theory with simulations, which was performed over 200 hours of experimental time. (a,d) Snapshots from simulations for high salt (high energy) and low salt (low energy) respectively, at the end of 200 hour time scale. (b,e) Mass fraction statistics from simulations for high and low energies respectively at the end of 200 hour timescale  compared to the predictions from equilibrium theory. Green dashed lines indicate the equilibrium theory predictions for mass-fraction of self-limiting trimers. (c,f) Time evolution of the mass-fraction of self-limiting trimers shown by magenta square markers, along with monomer mass-fraction in grey circle markers. The black filled stars denote the experimental measurements of T1 mass fraction at 4, 8 and 20 hours respectively in increasing order.}
\label{fig:Fig.6}
\end{figure*}

\jv{
By comparing the experimentally measured mass-fraction of aggregates with that obtained from the simulations, we obtain an estimate of the effective flat-flat interaction energies (see Supporting Information, \sref{sec:estimate_Wss}). \nref{fig:Fig.S9}(a) thus clearly shows that higher salt corresponds to an energy of approximately $6.5k_BT$, whereas the lower salt corresponds to a lower energy of $2.7k_BT$. Using these inferred interaction strengths, we performed simulations to reproduce the assembly process over a time scale of 4 hours. The resulting morphologies and corresponding mass fraction statistics at the end of 4 hours are presented in \nref{fig:Fig.5}(b,e) and (c,f), respectively, and shows good agreement with the experimental observations. The simulations are then extended over a time-scale of 200 hours to reach \ml{an assembly steady state}. Simulation snapshots of these two cases at the end of 200 hours are shown in \nref{fig:Fig.6}(a,d). The snapshots indicate a significant increase in the yield of self-limiting trimers in both cases. However, a considerable fraction of larger aggregates remains in the high-salt case (\nref{fig:Fig.6}(a)) at the end of the simulation.
}

\vspace{0.2cm}

\ml{
To understand this overabundance of large aggregates with respect to the more stable self-limiting T1 trimers in both simulations and experiments, we analytically predict the outcome of our aggregation process at equilibrium. We enumerate all possible aggregates compatible with the binding geometry of our hexagonal colloids, and compute their concentration at thermal equilibrium using an ideal gas of clusters formalism~\cite{israelachvili1976theory} (\sref{sec:equilibriumtheory}).}
\jv{
The mass fraction of the aggregates obtained at the end of 200 hours from the simulations are then compared with that obtained from \ml{the theory} (\nref{fig:Fig.6}(b,e)). The theory predicts a higher equilibrium yield of self-limiting trimers at 20~mM salt concentration than at 15~mM. Comparison with the simulation results at 200 h shows that the trimer yield at 20~mM salt remains below the equilibrium-theory prediction (\nref{fig:Fig.6}(b)), whereas at 15~mM the simulation results are in very good agreement with theory (\nref{fig:Fig.6}(e)). The noticeable difference between theory and simulation for higher salt could be due to the fact that the theory does not account for kinetic traps, which plays a stronger role at higher energies (higher salt) as opposed to lower energies (low salt) which allow for reversible bonds. The time evolution of the self-limiting trimers T1 in \nref{fig:Fig.6}(c,f) shows how kinetic traps affect the yield. For higher salt concentration (higher energy) the mass fraction of self-limiting trimers saturates around 10 hours, whereas for the lower salt concentration (lower energy) it keeps rising until it reaches the equilibrium value predicted from theory (shown by green dashed lines in \nref{fig:Fig.6}(b, c, e, f)), confirming that in the experiments, higher salt would allow for irreversible aggregates, whereas lowering the salt concentration allows for reversible bonds/error-correction and higher yield of self-limiting trimers with time.} \jh{Taken together, these observations shows that modulating the attraction energy between attractive sides in this colloidal system can be used to favor situations where the only stable configuration is the one where attractive bonds cooperate to stabilize the assembly.}

\section{Discussion} \label{sec:Discussion}

Our results demonstrate that depletion-induced interactions between colloidal particles can be made highly specific by controlling the surface topography of the colloids. 2D polygonal colloids with flat and corrugated sides interact preferentially through their flat faces, while interactions through the corrugated faces are largely suppressed, enabling the formation of aggregates with well defined topologies. The restriction of bonding to particular sides enables the formation of aggregates of limited size or predefined structure, providing a route to direct self-assembly outcomes purely through physical design rather than chemical modification. Similar forms of interaction selectivity have been reported in other colloidal systems, where surface roughness is used to localize depletion attractions to specific regions of a colloid, particularly when the depletant size is smaller than the characteristic roughness length scale \cite{zhao2007directing, badaire2008experimental, kraft2012surface}. In such cases, the polymer can partially penetrate surface asperities, diminishing excluded volume overlap and suppressing attractive interactions at roughened sides. However, in the present system the typical spatial periodicity of the corrugated surface is substantially larger than the size of the depletant, placing the system in a regime where complementary corrugated features could, in principle, interlock effectively, thereby increasing the excluded-volume overlap ($\Delta V$) and consequently generating stronger depletion attractions than those between flat sides. The absence of such interactions suggests that the suppression of depletion attraction cannot be attributed solely to the roughness length scale. We hypothesize that the suppression of depletion induced interactions at the corrugated sides arises from the precise surface architecture, which may limit the geometric complementarity between approaching corrugated surfaces.

\begin{figure}[h]
\centering
\includegraphics[scale=0.925]{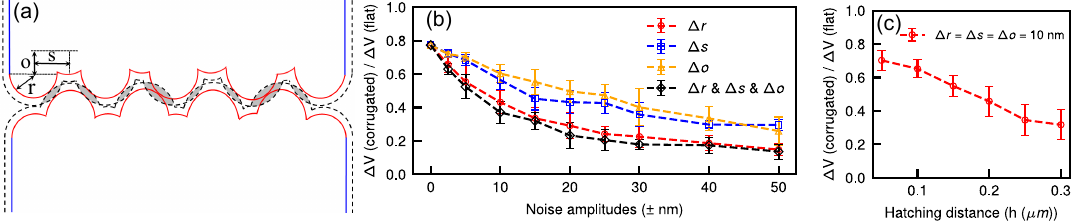}
\caption{(a) Schematic depiction of the wavy surface topography of the corrugated sides, parameterized by the radius of curvature ($r$), periodic spacing ($s$), and offset ($o$), each incorporating a normal perturbation of noise to mimic nanoscale roughness. (b) Plot showing the fractional overlap of excluded areas between two wavy surfaces to flat surfaces as a function of stochastic noise amplitudes in the individual parameters ($r$, $s$, and $o$) and their combined variation. (c) Progressive decrease of fractional overlap of excluded area of corrugated surfaces with increasing hatching distances at a fixed noise amplitude of 10 nm.}
\label{fig:Fig.7}
\end{figure}

To examine this quantitatively, we construct a simplified two-dimensional model of the corrugated lateral surface that captures the essential features observed in the SEM images of corrugated side (\nref{fig:Fig.1}(c)). The corrugated surface profile is parameterized using three geometric quantities: the voxel radius $r \approx 0.2\mu m$, the horizontal spacing between consecutive hatch lines $s \approx 0.3\mu m$, and the vertical offset between successive printed lines $o \approx 0.2\mu m$, as illustrated in \nref{fig:Fig.7}(a). One possible origin of these vertical offsets ($o$) is the bidirectional laser scanning process inherent to the two-photon polymerization technique. In this process, alternating scan directions introduce a vertical displacement between adjacent hatch layers, contributing to the observed surface roughness (\sref{sec:dir_hatching}). To capture depletion interactions within this framework, each surface is assigned with a penetrable region mimicking the excluded region corresponding to the effective size of the depletant. As two corrugated surfaces approach, the overlap between these regions is computed numerically, providing a quantitative measure of the depletion interaction. When the surface parameters are taken exactly as designed, the numerical model predicts a maximal overlap area between the exclusion regions, corresponding to a strong depletion-induced attraction as predicted before. We propose that in experiments, small deviations in the surface parameters resulting from the limits of printing precision or inherent fabrication noise could reduce the overlap and alter the effective interaction. To explore this possibility, Gaussian-distributed perturbations are independently applied to the voxel radius $r$, hatch spacing $s$, and vertical offset $o$ to mimic potential experimental variability. Among these parameters, fluctuations in $r$ produce the most pronounced reduction in excluded-volume overlap compared to variations in $s$ or $o$ (\nref{fig:Fig.7}(b)). 
Notably, nanometer-scale deviations of approximately \jh{10~nm}, corresponding to 5\% of the nominal voxel radius of 0.2~$\mu m$, lead to a substantial decrease in overlap volume compared to the flat-flat configuration, as evidenced by the progressive decline of fractional overlap volume shown in \nref{fig:Fig.7}(b). These results indicate that even small, experimentally plausible perturbations in surface geometry can significantly weaken the depletion attraction, providing an explanation for the suppression of interactions along the corrugated sides observed in experiments.


When similar perturbations are applied simultaneously to all three geometric parameters, possibly a close approximation to experimentally realistic surface roughness, the reduction in overlap volume follows the same trend and remains dominated by fluctuations in voxel radius. This effect is schematically illustrated in \nref{fig:Fig.7}(b), where a uniform perturbation of 10~nm across all parameters results in a two to threefold decrease in fractional overlap volume. These numerical findings are consistent with the experimental observation that corrugated sides do not participate in depletion bonding. The same mechanism also explains the absence of significant interaction between flat and corrugated sides, where mismatches in local geometry hinder the formation of sustained excluded volume overlap and effectively suppress depletion-induced attraction. 

Nevertheless, interactions involving the corrugated surfaces become increasingly relevant when the hatching distance is reduced and when the overall interaction strength is enhanced by adjusting the concentrations of the depletant and salt (\nref{fig:Fig.3}). Decreasing the hatching distance effectively reduces the amplitude of the surface corrugation, rendering the lateral faces progressively flatter and thereby more compatible with depletion-induced contacts (\nref{fig:Fig.3}(a-e)). Under these conditions, the geometric mismatch that suppresses interactions at larger hatching distances is partially alleviated, allowing corrugated sides to participate in bonding (\nref{fig:Fig.3}(b)). Numerical calculation also predicts the consistent decrease in excluded overlap volume with increasing hatching distance $h$ at a fixed noise amplitude of 10 nm in all surface constructing geometric quantities(\nref{fig:Fig.7}(c)) . In parallel, increasing the interaction strength amplifies the contribution of even relatively small excluded-volume overlaps. At sufficiently high depletant concentrations, these modest overlaps can provide enough stabilization to sustain contacts involving corrugated surfaces. As a consequence, interactions that are otherwise unfavourable become accessible, leading to a loss of directional selectivity (\nref{fig:Fig.3}(c,d)). Together, these observations explain why specific interactions are observed only within a well-defined window of design and solution parameters, and highlight the balance between surface geometry and interaction strength required to maintain selective depletion-driven assembly.

\section{Conclusion}

In conclusion, we have demonstrated that surface morphology engineered through two-photon polymerization provides a powerful means of regulating depletion-mediated interactions between colloids. By controlling the corrugation and spatial distribution of interacting sides, we encoded directional interaction rules that guided the formation of targeted assemblies ranging from simple dimers to complex two-dimensional structures. Our results further show that interaction specificity emerges from a delicate balance between particle design and interaction strength, with both surface architecture and solution parameters playing critical roles in determining assembly outcomes. More broadly, this work establishes surface topography as a versatile design parameter for programming colloidal interactions and self-assembly. The ability to translate fabrication-defined surface features into predictable collective behavior offers new opportunities for the rational design of colloidal building blocks and the exploration of emergent organization in soft-matter systems.

\section{Experimental Section}
\subsection*{Fabrication of colloids}

Two-photon polymerization technique was employed to fabricate 2D colloidal particles with precise and customizable shapes \cite{tigges2016hierarchical,mayarani2025lifetime}. A thin ($\sim$170 $\mu$m) circular coverslip (diameter = 30~mm) was used as the substrate for fabrication. The coverslip was thoroughly rinsed sequentially with acetone and isopropanol, followed by treatment with oxygen plasma. A sacrificial layer ($\simeq$35~nm) of polyacrylic acid (PAA, 2 wt$\%$) was deposited via spin coating at 3000~rpm for 20 seconds. Subsequently, a drop of resin (IP-L 780, Nanoscribe, GmbH) was dropcast onto the coated side, while a drop of immersion oil was applied to the opposite side. The coverslip was then loaded into the printer operating in conventional mode where the Laser is focused on to the resin through the immersion oil. The focal plane of the laser is set at the resin-substrate interface. A three-dimensional (3D) model of the desired particle is designed using computer-aided design software (CATIA V5) and exported in STL format. The file is subsequently imported into the printing software (NanoWrite), which directs the laser writing process based on the predefined design parameters. Upon completion of the printing, the substrate was immersed in a developer solution of propylene glycol monomethyl ether acetate (PGMEA) for 20 minutes to remove unpolymerized resin. Importantly, PGMEA does not noticeably degrade the underlying PAA sacrificial layer, helping to maintain adhesion of the printed structures to the substrate.

\subsection*{Preparation of an experimental chamber}

Studying colloidal self-assembly requires a closed chamber to prevent evaporation. A circular hole of diameter 4 mm was made in double-sided tape (thickness = 0.2~mm) to form a chamber around the printed colloids. Then, approximately 3-4~$\mu$l of depletant solution containing PEG as depletant, Tergitol NP-10 as surfactant, and salt was gently added. The pH of the solution is controlled at 6.0 by using a 5 mM MES hydrate buffer titrated with NaOH. This solution spontaneously dissolved the PAA sacrificial layer, enabling the colloids to levitate and exhibit Brownian motion. To prevent leakage, a thin hydrophobic barrier was applied around the chamber perimeter using a hydrophobic pen. The chamber was subsequently sealed with another coverslip. A 63X objective lens was employed to focus on the colloids, and the resulting images and videos were captured using a digital camera (MIchrome 6). Further analysis was conducted using ImageJ software alongside custom-written Python scripts.

\subsection*{Numerical simulations}

Brownian dynamics simulations were used to rationalize the experimentally observed formation of self-limiting trimers and to distinguish near-equilibrium assembly from kinetically trapped structures. The hexagonal colloids were represented as rigid  particles with two consecutive attractive sides and repulsive hard-core interactions on the remaining sides, following the experimental colloid design. Interaction strengths were varied in units of thermal energy to compare different binding regimes, while cluster populations were analyzed from the simulated trajectories using the same trimer and aggregate classifications applied to the experiments. Additional details of the model geometry, interaction potentials, simulation protocol, are given in the Supporting Information.

\subsection*{Supporting Information} \par 
Supporting Information is attached below.

\subsection*{Data availability}
The codes for both MD simulation and numerical calculation can be accessed at: \url{https://codeberg.org/jvishnu/Corrugated_Surfaces_numericals}. The data that support the findings of this study will be made available upon reasonable request.


\subsection*{Acknowledgements} \par 

This work was supported by ANR Grant No. ANR-22-CE30-0024-01 and the Impulscience® program from Fondation Bettencourt-Schueller. M.L. was supported by ANR Grant No. ANR-22-ERCC-0004-01 as well as ERC Starting Grant No. 677532. M.L., OdR and JH belong to the CNRS consortium AQV. This work used HPC resources from GENCI-IDRIS (Grant 2026-AD010918055), as well as computational resources from the ``M\'esocentre'' computing center of Universit\'e Paris-Saclay, CentraleSup\'elec and \'Ecole Normale Sup\'erieure Paris-Saclay supported by CNRS and R\'egion \^Ile-de-France (\url{https://mesocentre.universite-paris-saclay.fr/}). The 3D printer setup was acquired through SESAME Region Ile de France MiLaMiFab grant. We thank Justine Laurent for her valuable assistance and guidance in operating the 3D printing system.

\newpage


\renewcommand{\thefigure}{S.\arabic{figure}}
\renewcommand{\thesection}{S\arabic{section}}
\renewcommand{\thesubsection}{S\arabic{section}.\arabic{subsection}}
\renewcommand{\thetable}{S\arabic{table}}

\setcounter{table}{0}
\setcounter{figure}{0}
\setcounter{section}{0}
\setcounter{subsection}{0}
\setcounter{equation}{0}
\begin{center}
{\textbf{\Large Supplementary information}}\\
\vspace{0.4cm}
\textbf{ \Large Designing corrugated surfaces to guide colloidal self-assembly}\\
\vspace{0.4cm}

Dinesh Kumar Sahu$^{1,2}$*, Jude Ann Vishnu$^{1}$, Lisa Shafroth$^{2}$, Martin Lenz$^{1,2}$, Olivia du Roure$^{2}$ and Julien Heuvingh$^{2}$*\\

$^{1}$Universit\'e Paris-Saclay, CNRS, LPTMS, 91405, Orsay, France\\

$^{2}$PMMH, CNRS, ESPCI Paris, PSL University, Sorbonne Universit\'e, Universit\'e Paris-Cit\'e, 75005, Paris, France\\

$^{•}$*Emails : dinesh-kumar.sahu@espci.fr, julien.heuvingh@u-pariscite.fr

\end{center}

\section{Fabrication of colloids} \label{sec:fabrication}

\subsection{Experimental setup}

\begin{figure}[h]
\centering
\includegraphics[scale=0.925]{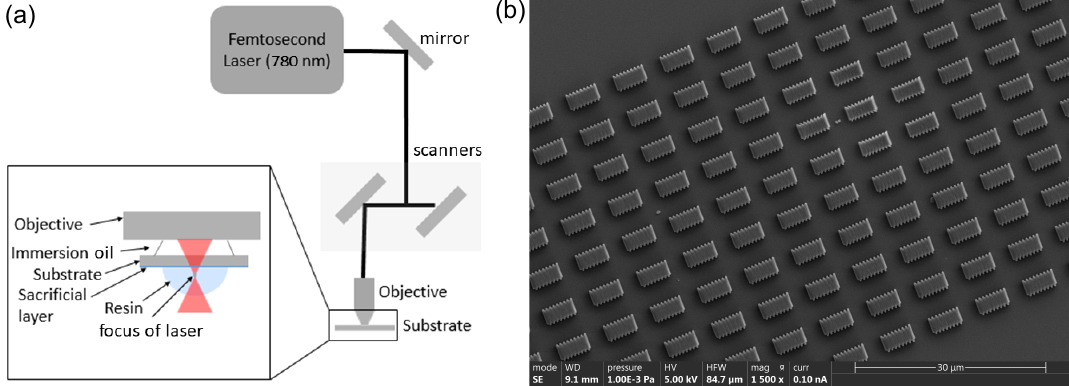}
\caption{(a) Schematic depiction of two-photon polymerization technique in conventional mode. (b) SEM image of rectangular array of printed colloids.}
\label{fig:Fig.S1}
\end{figure}

Colloidal particles were fabricated using high-resolution two photon polymerization on a Nanoscribe Photonic Professional GT2 system (Nanoscribe GmbH, Germany). All structures were written in conventional mode, as decribed in the schematic (\nref{fig:Fig.S1}(a)). The nonlinear absorption intrinsic to two photon polymerization confines polymerization to the ellipsoidal focal volume, referred to as a voxel, enabling precise and high fidelity reproduction of the designed three dimensional geometries \cite{bunea2021micro, pagliano20233d}.

Laser exposure conditions were controlled by adjusting the average laser power, and the scan speed, fixed at 10 mms$^{-1}$. These parameters collectively determine the effective voxel dimensions through the polymerization threshold of the resist. Lateral positioning of the laser focal volume was achieved using high speed galvanometric mirrors, while axial motion was controlled by a piezoelectric stage with nanometer scale precision. This combination allowed accurate implementation of user defined slicing, hatching, and contour writing parameters. Directional hatching, slicing, and contour writing strategies were employed to intentionally modulate the lateral surface morphology of the colloids, thereby enabling selective depletion mediated interactions in subsequent assembly experiments. Colloids were typically fabricated in rectangular arrays, with a minimum side-to-side separation of 2 $\upmu$m between adjacent particles to prevent unintended fusion during writing and development. A representative scanning electron microscopy image of several rectangular colloids fabricated using this protocol is shown in \nref{fig:Fig.S1}(b).

\subsection{Hatching strategy for printing colloids}\label{sec:hatching_strategy}

\begin{figure}[h]
\centering
\includegraphics[scale=0.925]{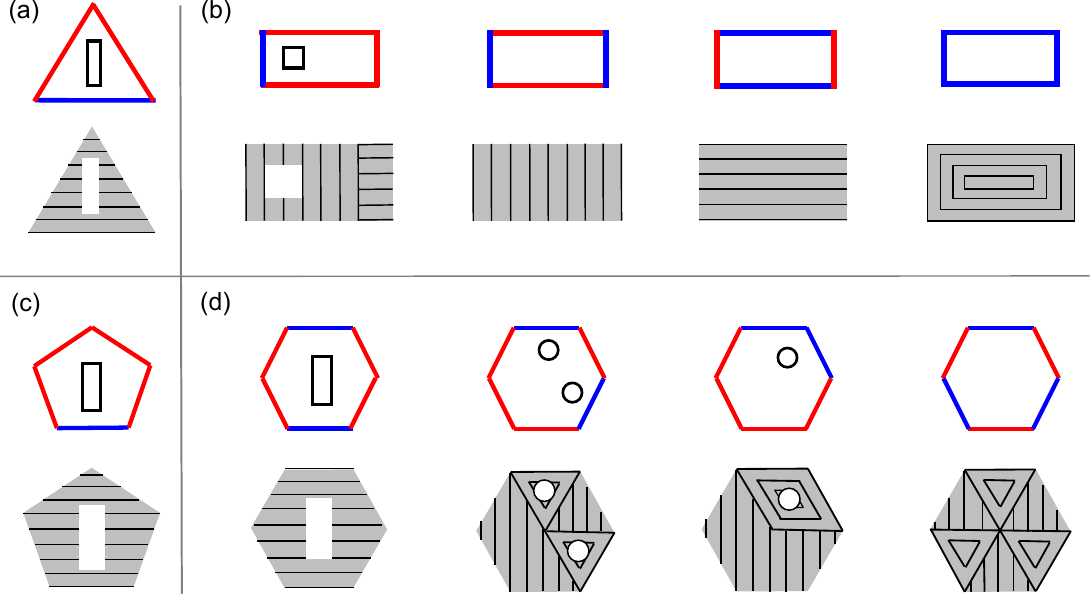}
\caption{Schematic (top) and corresponding hatching strategy (below) of (a) triangular, (b) rectangular, (c) pentagonal and (d) hexagonal colloids used for fabrication.}
\label{fig:Fig.S2}
\end{figure}

The specificity of colloidal interactions in this system is governed primarily by the engineered architecture of the colloid surfaces. Here, we exploit three dimensional laser writing parameters to deliberately tailor the lateral surfaces of the colloids to be either flat (favourable) or corrugated (unfavourable) for depletion mediated interactions. In the schematics shown in the main text, sides designated as favourable are indicated in blue and are fabricated using continuous laser scanning, whereas unfavourable sides, shown in red, are produced using discrete laser scanning defined by a hatching distance of 0.3 $\mu m$. As an illustrative example, in \nref{fig:Fig.S2}(a-c), the hatch lines are oriented parallel to the bottom side of a triangular colloid, indicated in blue in the schematic, resulting in a smooth lateral surface that preferentially supports depletion attraction. The remaining sides, fabricated with non parallel hatching, exhibit corrugated surface textures and are thus rendered unfavourable for interaction. In order to make every side favourable for interaction, the hatching lines are simply replaced by contour lines as shown for rectangular colloids in \nref{fig:Fig.S2}(b).

Fabrication of colloids with multiple favourable sides requires part wise printing, in which different regions of the same colloid are written using distinct scanning strategies. In this approach, favourable sides are printed using contour writing, while unfavourable sides are generated using perpendicular or inclined hatching, as demonstrated for hexagonal colloids in \nref{fig:Fig.S2}(d). Printing parameters for contour and hatching are selected to preserve voxel overlap and structural integrity, while enabling controlled variation of surface morphology across different sides.

\section{Diffusion of colloids} \label{sec:diffusion}

\begin{figure}[h]
\centering
\includegraphics[scale=0.925]{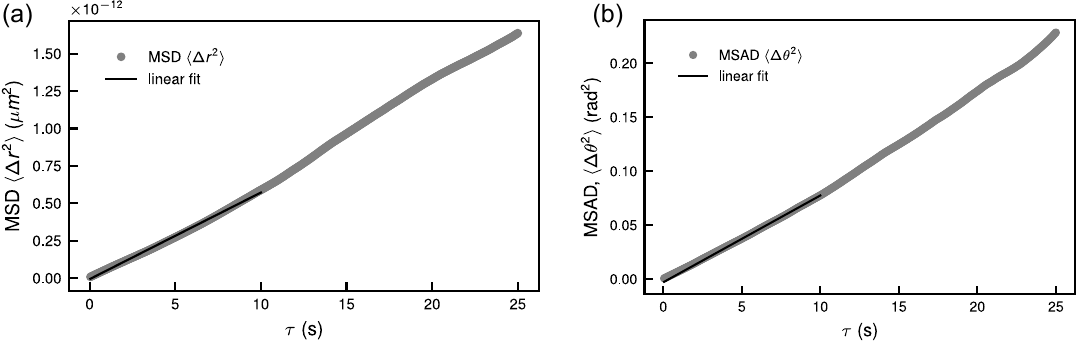}
\caption{(a) Mean square displacement (MSD) and mean square angular displacement (MSAD) vs lag time for rectangular colloids in a depletant solution of 0.01 mg/ml of PEG-0.6MDa. Straight black lines are the linear fits.}
\label{fig:Fig.S3}
\end{figure}

Upon dissolution of the poly(acrylic acid) sacrificial layer, the rectangular colloids are released from the substrate and undergo thermally driven Brownian diffusion, predominantly confined to two dimensional motion parallel to the substrate. The system is allowed to equilibrate for 10 to 15 minutes to ensure that the colloids reach stable sedimentation heights before measurements are performed. Colloid motion is then recorded and the centroid positions of individual colloids are tracked using ImageJ. The resulting time series of colloid coordinates are used to compute the mean squared displacement (MSD) according to the following equation.
\begin{equation}
<\Delta r^{2}> = 4D_{t}\tau
\end{equation}
\noindent where $D_{t}$is the translational diffusion coefficient and $\tau$ is the time lag. The analysis was performed for multiple colloids, and the resulting average mean squared displacement $<\Delta r^{2}>$ is shown in \nref{fig:Fig.S3}(a). From the linear regime of the MSD, the translational diffusion coefficient was determined to be 0.033$\pm$0.002 $\upmu$m$^{2}s^{-1}$. For comparison, we also estimated the translational diffusion coefficient theoretically. Using the Stokes Einstein relation, the diffusion coefficient of an equivalent sphere with the same volume as the rectangular colloid is calculated to be approximately 0.13 $\upmu$m$^{2}s^{-1}$. Accounting for the effects of particle anisotropy \cite{allison1999low} and hydrodynamic interactions arising from the close proximity of the colloid to the substrate \cite{faucheux1994confined}, the predicted translational diffusion coefficient is reduced to approximately 0.04 $\upmu$m$^{2}s^{-1}$, which is in good agreement with the experimentally measured value.

\vspace{0.2cm}

\noindent In a similar manner, the rotational dynamics of the colloids were quantified by tracking their angular orientations over time. The orientation of each colloid was determined by fitting an ellipse to its shape in ImageJ, which provides the angle between the major axis of the colloid and the x-axis, allowing the time-dependent angular position $\theta(t)$ to be extracted for individual trajectories. From the resulting angular time series, the mean squared angular displacement was calculated as
\begin{equation}
<\Delta \theta^{2}> = 2D_{r}\tau
\end{equation}
\noindent where $D_{r}$ is the rotational diffusion coefficient. The analysis was performed for multiple particles, and the average mean squared angular displacement exhibited a linear dependence on $\tau$ over the measured time window (\nref{fig:Fig.S3}(b)). From the slope of this linear regime, the rotational diffusion coefficient was determined to be 0.004 rad$^{2}$s$^{-1}$.

\section{Directional laser scanning affects interaction specificity} \label{sec:dir_hatching}

\begin{figure}[h!]
\centering
\includegraphics[scale=0.925]{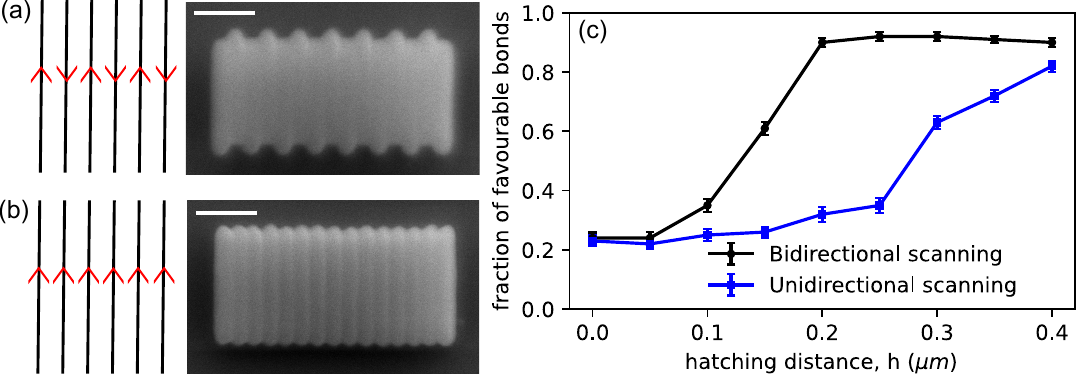}
\caption{ Bidirectional (a) and unidirectional (b) scanning of laser along the consecutive hatching lines and the corresponding SEM images of printed colloids. (c) Difference in fraction of favourable contacts as a measure of interaction specificity between unidirectional and bidirectional scanning with different hatching distances.}
\label{fig:Fig.S4}
\end{figure}

We observe enhanced interaction specificity for rectangular colloids with favourable short sides when fabricated using bidirectional laser scanning (schematically shown in \nref{fig:Fig.S4}(a)) at hatching distances greater than 0.2 $\upmu$m, as shown in \nref{fig:Fig.S4}(c). Here we also have the freedom of unidirectional Laser writing along the hatch lines. A direct comparison with unidirectional scanning reveals pronounced differences in the resulting surface morphology, which are clearly visible in the corresponding SEM images (\nref{fig:Fig.S4}(a,b)). In particular, bidirectional scanning produces a corrugated surface with a wavelength, defined as the distance between successive crests or troughs, that is approximately twice that obtained with unidirectional scanning. This increase in wavelength effectively reduces the local curvature of the corrugated surface, rendering the longer sides comparatively flatter and therefore less favourable for depletion mediated interactions. In contrast, unidirectional scanning generates a finer corrugation pattern, which requires a larger hatching distance around 0.4 $\upmu$m, to sufficiently suppress interactions along the longer sides and achieve a high fraction of favourable bonds (\nref{fig:Fig.S4}(c)). It should be noted that hatching distances exceeding approximately 0.4 $\upmu$m cannot be reliably employed, as insufficient voxel overlap leads to poor structural continuity and deformation of the printed colloids.

\section{Numerical calculation of overlap excluded area}

\begin{figure}[h]
\centering
\includegraphics[scale=0.925]{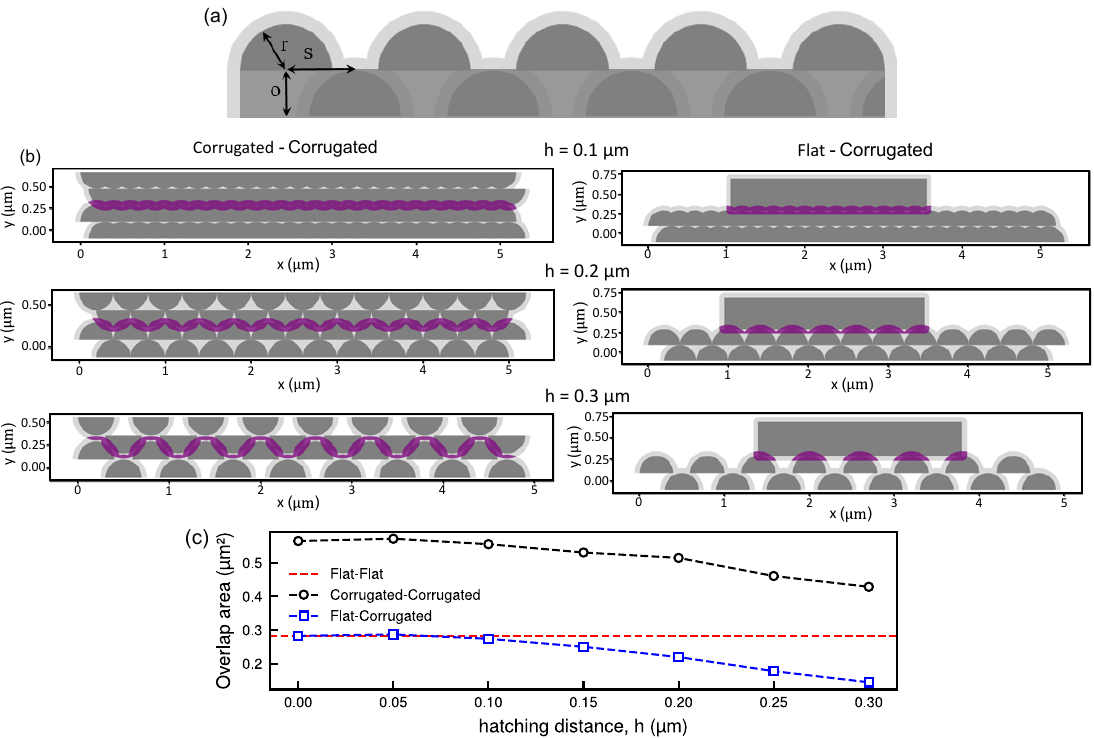}
\caption{(a) Schematic representation of the two dimensional geometric model of a corrugated surface constructed using the surface parameters $r$, $s$ and $o$. (b) Representative snapshots of the numerical configurations corresponding to the maximum excluded area overlap for corrugated-corrugated and flat-corrugated surface pairs at different hatching distances. (c) Dependence of the excluded area overlap  on the hatching distance $h$ for corrugated corrugated (black circles) and flat corrugated (blue squares) configurations. The red horizontal dotted line indicates the excluded area overlap for two flat surfaces of length equal to half that of the corrugated surface, mimicking a rectangular colloid used in the experiments.}
\label{fig:Fig.S5}
\end{figure}

We anticipate that the specificity of depletion mediated interactions observed in this system originates from the precise topography of the corrugated lateral surfaces, which limits the effective overlap of excluded volume between two such sides. In this picture, the corrugated geometry substantially reduces the entropic gain associated with depletion attraction, thereby rendering these surfaces unfavourable for sustained particle binding. To quantitatively evaluate this effect, we performed numerical calculations to estimate the excluded area overlap generated when two corrugated surfaces are brought into face to face contact. For simplicity and computational tractability, the corrugated surface was approximated as a periodic series of semicircular protrusions with radius equal to the voxel radius $r \sim 0.2 \mu$m. The lateral positions of these semicircles were defined by the hatch spacing $s \sim 0.3 \mu$m, while the vertical offsets between adjacent rows were introduced using the parameter offset $o \sim 0.2 \mu$m, thereby reproducing the experimentally observed wavy surface morphology (\nref{fig:Fig.7}(a) and \nref{fig:Fig.S5}(a)). The corrugated surface profile was generated using the Python package shapely, which allows for precise construction and manipulation of complex planar geometries.

To model excluded part, each semicircular feature was surrounded by a penetrable exclusion region with a uniform thickness of $0.056 \mu$m, corresponding to the radius of gyration of PEG 0.6MDa. This exclusion region represents the volume inaccessible to polymers near the colloid surface and thus defines the local excluded area relevant for depletion attraction. When two such corrugated surfaces are brought into contact, the degree of overlap between their respective exclusion regions provides a direct measure of the entropic gain available to the system upon colloid association. A representative schematic of the constructed geometry and the associated exclusion regions is shown in \nref{fig:Fig.S5}(a). In the numerical procedure, one corrugated surface was fixed at the bottom, while a second, identical surface was translated vertically from above until first contact was established. The upper surface was then systematically scanned along the horizontal direction to identify the configuration that maximized the overlap between the respective exclusion regions. This procedure was repeated for different values of the hatch spacing parameter $s$, which directly determines the periodicity of the corrugated surface. Representative snapshots of the configurations corresponding to maximum excluded area overlap for several values of $s$ are shown in \nref{fig:Fig.S5}(b), where the overlapped regions are highlighted in violet. A clear reduction in the overlap area is observed with increasing $s$, indicating a progressive weakening of depletion mediated attraction between corrugated faces. This trend is consistent with experimental observations, where larger hatching distances lead to enhanced interaction specificity by suppressing bonding along corrugated surfaces, as quantified in \nref{fig:Fig.S5}(c). An analogous analysis was performed for interactions between a flat surface and a corrugated surface, where the length of the flat surface is taken half of the corrugated surface in order to mimic the rectangular shape of the colloids used in the experiment. In this case, the calculated overlap area was consistently smaller than that obtained for the flat-flat configuration, particularly at larger values of $h$ or equivalently larger $s$ (\nref{fig:Fig.S5}(b,c)). This reduction in excluded area overlap provides a direct explanation for the experimentally observed absence of bonding between flat and corrugated sides, as the entropic driving force for depletion attraction in such configurations is insufficient to stabilize particle contacts.

\begin{figure}[h]
\centering
\includegraphics[scale=0.925]{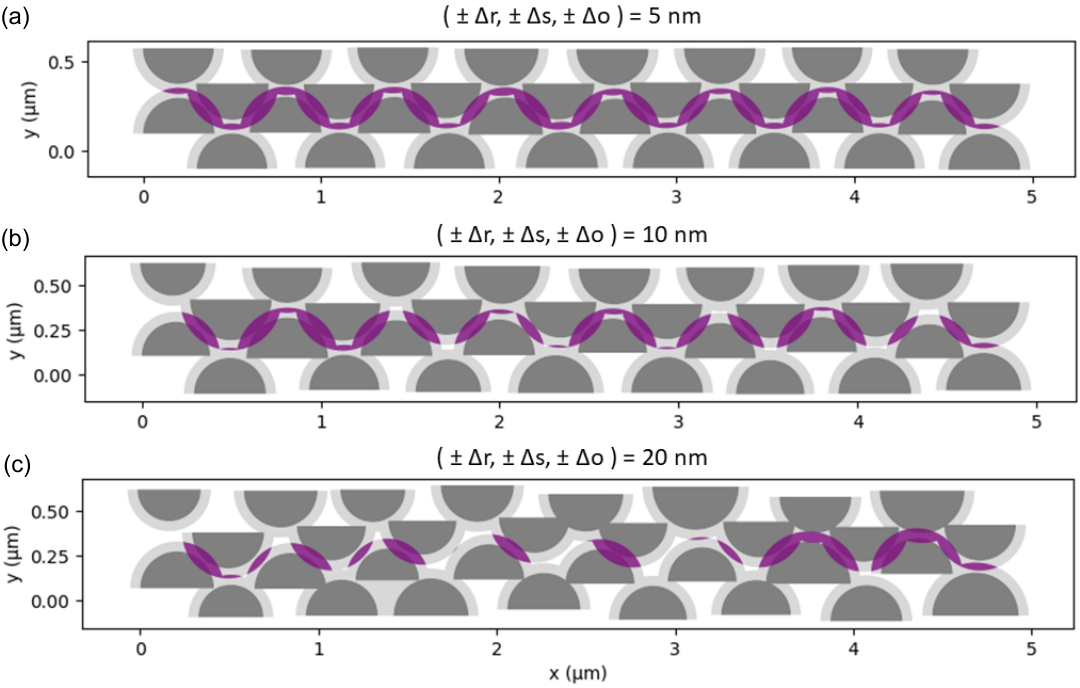}
\caption{Representative snapshots from single numerical trials showing the configurations corresponding to maximum excluded area overlap between two opposing corrugated surfaces. Independent Gaussian distributed noise was introduced into the geometric parameters with amplitudes of (a) 5 nm, (b) 10 nm, and (c) 20 nm. Increasing noise amplitude leads to a progressive reduction in the overlap area.}
\label{fig:Fig.S6}
\end{figure}

We first note that for an idealized corrugated surface constructed with parameters satisfying conditions for a hatching distance of $h = 0.3 \upmu$m, the calculated excluded area overlap between two opposing corrugated sides is larger than that obtained for two flat sides (\nref{fig:Fig.S5}(c)). This result would, in principle, suggest that corrugated sides should experience stronger depletion attraction than flat sides. However, this prediction directly contradicts our experimental observations, which consistently show that depletion mediated interactions are significantly more favourable between flat sides than between corrugated ones. This discrepancy indicates that additional factors not captured in the idealized geometric model must play a role in suppressing interactions along corrugated sides. We therefore consider the effect of small deviations in the geometric parameters used to construct the corrugated surface. Such deviations are naturally expected to arise from finite printing precision or minor fluctuations in laser exposure during fabrication.
\vspace{0.1cm}

To examine this effect, we carried out numerical calculations in which independent Gaussian distributed perturbations were applied to each surface defining parameter, namely the voxel radius $r$, the hatch spacing $s$, and the vertical offset $o$, centered around their respective mean values of 0.2 $\mu m$, 0.3 $\mu m$, and 0.2 $\mu m$. Representative snapshots from single simulation trials at increasing noise amplitudes ($\pm \Delta r, \pm \Delta s, \pm \Delta o$) are shown in \nref{fig:Fig.S6}(a–c). These results clearly demonstrate a sharp reduction in the excluded area overlap compared to the noiseless case, along with a monotonic decrease in overlap as the noise amplitude increases(\nref{fig:Fig.7}(b)). For $h = 0.3 \mu$m, the excluded area overlap between a flat and a corrugated side is already smaller than that between two flat sides(\nref{fig:Fig.S5}(c)). The introduction of geometric noise further reduces this overlap, thereby weakening the depletion attraction in flat-corrugated contacts even more. This combined effect provides a consistent explanation for the experimentally observed high interaction specificity, wherein flat sides dominate colloid binding while corrugated sides remain effectively non interacting.

\newpage

\begin{figure*}[h!]
    \centering
    \includegraphics[scale=0.925]{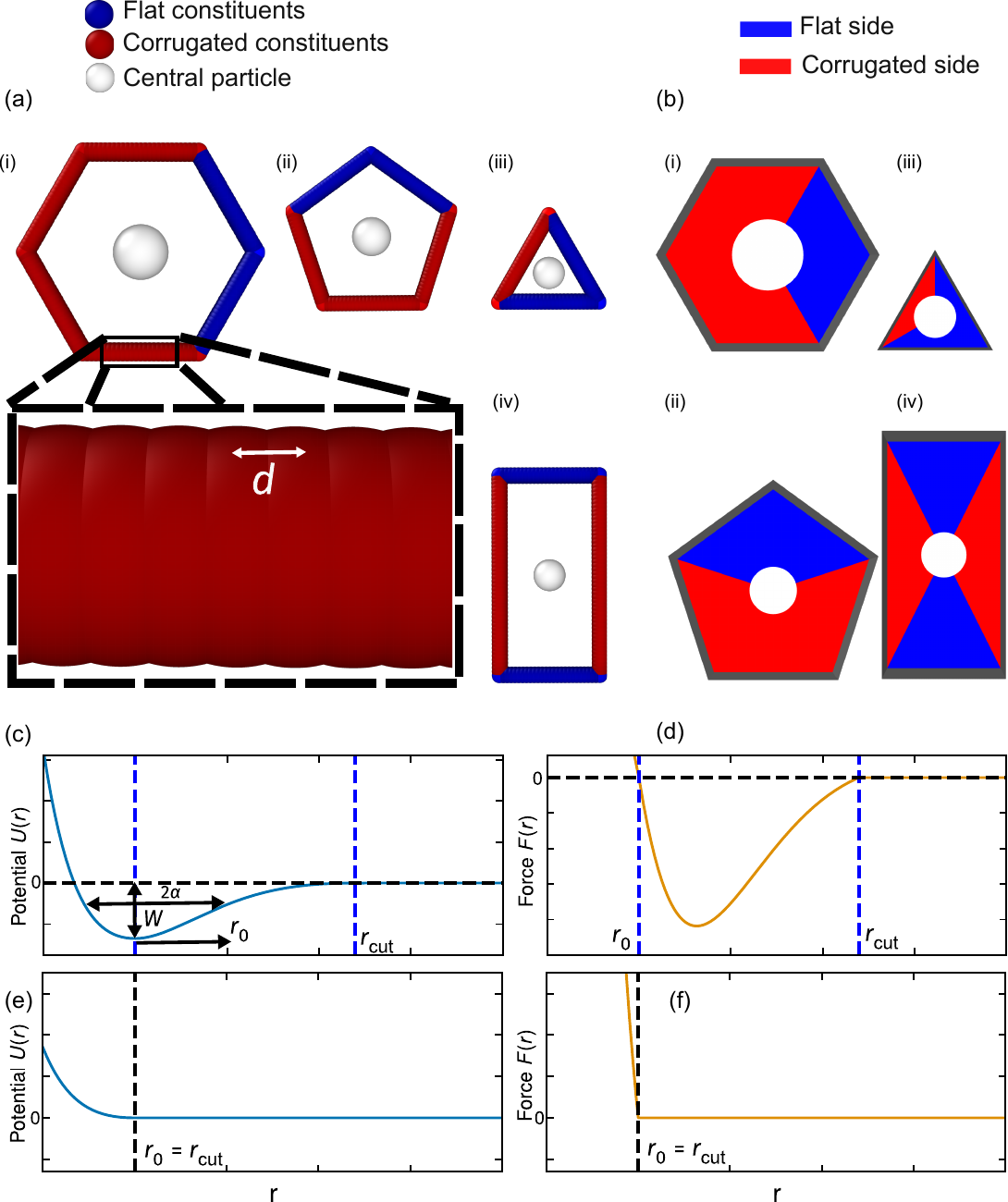}
    \caption{Simulation models and interaction potentials: (a) Different rigid-body geometries of the polygonal colloids, including (i) hexagons, (ii) pentagons, (iii) triangle, (iv) rectangles constructed from constituents particles. In each case, attractive edges (flat constituents) are shown in dark blue and repulsive edges ((corrugated constituents) in dark red. 
(b) Corresponding mesh representations of the polygonal particles, shown for clearer visualization and comparison with experiments. Red faces are repulsive/ corrugated, whereas blue faces are attractive/ flat. (c) Morse potential in XPLOR mode used in the simulations, with the potential parameters indicated in the plot.
(d) Force corresponding to the XPLOR-mode Morse potential. (e) Shifted Morse potential used in the simulations. (f) Force corresponding to the shifted Morse potential in (e)}
\label{fig:Fig.S7}
\end{figure*}
\section{Model and simulation details}\label{sec:model and sim}
In this section, we discuss the simulation model used. We start by describing the general construction of a 2D polygonal colloid in our simulation and the nature of interaction it can have in \sref{sec:poly_model}. Then in \sref{sec:interaction}, we describe the pair potential used for our colloidal model. Finally in \sref{sec:units}, we describe the units in ours simulation and how it relates to physical unit.
\subsection{Simulation model for polygonal colloids} \label{sec:poly_model}
The two dimensional colloids are modeled and simulated using the simulation package HOOMD-blue version 6.00 \cite{anderson2020hoomd}. Colloids are modeled as rigid body composed of a collection of beads (constituent particles) arranged to represent the desired polygonal geometry around a central particle (see Figure \ref{fig:Fig.S7} (a)). The constituent particles are placed along each side of the polygon at a constant spacing $d$, and are fixed relative to one another using HOOMD's rigid body  \cite{nguyen2011rigid, glaser2020pressure} framework, ensuring that the colloids underwent only translational and rotational motion in the $x$–$y$ plane. Distinct sets of constituent particles are assigned to different sides of the polygon, allowing specific sides to be designated as attractive, or having hard-core repulsion. To have correspondence with experiments, we only have either attractive or purely repulsive interactions. Constituents on the flat sides (dark blue edges/blue faces in \nref{fig:Fig.S7}(a,b)) interacts with other flat side constituents attractively, whereas corrugated constituents (dark red edges/red faces in \nref{fig:Fig.S7}(a,b)) interact with other corrugated / flat side constituents via hard-core repulsion. Visualization of particle geometries in \nref{fig:Fig.S7}(a-b))  are done with the help of OVITO \cite{stukowski2010visualization}. 

\subsection{Interaction Potential} \label{sec:interaction}

In the simulations, interactions are restricted to constituent particles belonging to different colloids, meaning that the intra-colloidal interactions are turned off.  Both attractive and repulsive hard-core inter-colloidal interactions between constituents are based on a pairwise Morse potential (see \neqref{eq:Morse}). 

\begin{align}\label{eq:Morse}
U_{\mathrm{Morse}}(r)&=
\begin{cases}
\mathcal{W} \left( e^{-2\alpha(r-r_0)} - 2e^{-\alpha(r-r_0)} \right), 
& \quad r < r_{\mathrm{cut}}, \\
0, & \quad r \ge r_{\mathrm{cut}} .
\end{cases}
\end{align}
Here $r_0$, $\alpha$, and $\mathcal{W}$ denote the position of the potential minimum, the interaction width, and the interaction strength, respectively for a constituent particle. To ensure finite interaction range, the potential is truncated at a cutoff distance $r_{\mathrm{cut}}$. Two distinct truncation schemes are employed for attractive and repulsive interactions. For attractive interactions, we employ an XPLOR-smoothed Morse potential. In this scheme, the potential in \neqref{eq:Morse} is multiplied by a polynomial switching function $S(r)$ (see \neqref{eq:smooth}) between an onset distance $r_{\text{on}}$ and a cutoff distance $r_{\text{cut}}$, such that both the potential and its first derivative vanish continuously at the cutoff. See \nref{fig:Fig.S7}(c)-(d) for the schematic of this potential and the corresponding force.

\begin{align}\label{eq:smooth}
S(r)&=
\begin{cases}
1, \quad  & r<r_{\mathrm{on}}, \\ \\
 \frac{(r^2_{\text{cut}}-r^2)^2(r^2_{\text{cut}}+2r^2 - 3r^2_{\text{on}})}{(r^2_{\text{cut}} - r^2_{\text{on}})^3}, \quad & r_{\mathrm{on}} \leq r \leq r_{\mathrm{cut}}, \\ \\
 0, \quad & r>r_{\mathrm{cut}}
\end{cases}
\end{align}
For repulsive interactions, a shifted Morse potential (see \neqref{eq:shift}) is used 
\begin{align}\label{eq:shift}
U_{\text{Morse shift}} &=
\begin{cases}
 U_{\mathrm{Morse}}(r) - U_{\mathrm{Morse}}(r_{\mathrm{cut}}),\qquad & r < r_{\mathrm{cut}}, \\ \\
0, \qquad & r \geq r_{\mathrm{cut}}, \\ \\
\end{cases}
\end{align}
such that the potential energy vanishes at the cutoff, $U_{\mathrm{Morse \ shift}}(r_{\mathrm{cut}}) = 0$. This construction yields a purely repulsive excluded-volume interaction, while retaining the short-range steepness of the Morse potential.  See \nref{fig:Fig.S7}(e)-(f) for the schematic of this potential and the corresponding force.

For a given pair of interacting side types $m$ and $n$, the attractive interaction is distributed over the constituent particles used to discretize each side. In the current model, which can be generalized to any 2D polygonal colloids, different sides may contain different number of constituent particles depending on whether the shape is regular/ irregular or because interacting sides have different physical lengths. Without an appropriate normalization, the total attraction between two sides would then depend artificially on the discretization, or on which of the two sides contains more constituent particles, rather than on the intended side–side interaction strength.
To avoid this, the interaction strength $\mathcal{W}$ between individual constituent particles on side type $m$ and $n$ is chosen as
\begin{align}
	\mathcal{W} &= 0.96 \ \frac{W_{m,n}}{\mathrm{min}(N_m,N_n)}
\end{align}
where $N_m$ and $N_n$ are the numbers of constituent particles used to represent the two sides and  $W_{m,n}$ denote the target effective side–side interaction strength in the perfectly aligned bound configuration. The numerical prefactor $0.96$ is determined from zero-temperature tests in which pairs of colloids are held at fixed separations, so that the resulting effective side–side attraction matches the desired target strength $W_{m,n}$ between the sides. This choice ensures that the total effective attraction is controlled by the number of constituent particles that can form one-to-one contacts across the interface, and therefore remains approximately independent of the discretization, including when the two interacting sides have unequal lengths.


\begin{table*}[h]
\centering
\caption{Simulation parameters used in the Brownian dynamics simulations of a hexagonal colloid}
\label{tab:sim_params}
\begin{tabular}{llll}
\hline
\hline
Parameter & Symbol & Hexagon \\
\hline
Temperature & $k_B T$ & $1$  \\[4pt]

Translational diffusion constant  & $D_t^{\text{sim}}$ & $0.1$ \\


Time step  &   $\Delta t_{\text{sim}}$  & $10^{-5}$ \\[4pt]

Rotational diffusion  constant  & $D_r^{\text{sim}}$ & $0.097$  \\[4pt]

Morse width parameter & $\alpha$ & $20.9$ \\[4pt]

Morse minimum position & $r_0$ & $0.4$ \\[4pt]

Cutoff radius & $r_{\text{cut}}$ & $0.52$ \\[4pt]

Constituent spacing & $d$ &  $0.096$ \\[4pt]


\hline
\hline
\end{tabular}
\end{table*}

\subsection{Simulation and Physical units} \label{sec:units}
All simulations are performed using a Brownian dynamic integrator. Thus the particle motion is diffusive and the inertial effects are neglected. Units corresponding to each quantity is expressed in simulation units, i.e., an internally consistent unit system in which length, energy, and time are defined by the numerical parameters of the model. Particle geometries and interaction distances are given directly in these units; for example, polygon shapes, side lengths, and interaction ranges are all specified within the same simulation length scale.The simulation energy unit is chosen such that $k_BT=1$, and interaction strengths are therefore expressed relative to the thermal energy. Time is expressed in simulation units defined by the integration timestep $\Delta \ t_{\mathrm{sim}}$. All other simulation parameters are listed in Table \ref{tab:sim_params}. 

To relate simulation results to experiment, simulation quantities are mapped to physical units using the experimental particle geometry and translational diffusion coefficient. The simulation length scale is chosen such that one simulation length unit corresponds to $1\mu m$, allowing particle dimensions and interaction ranges to be specified directly in physical length units. For the purpose of converting simulation time to physical time, we further define a characteristic particle length $L_0 = \sqrt{A_{\mathrm{poly}}}$, where $A_{\mathrm{poly}}$ is the projected area of the polygon. This quantity is used only in the diffusion based time mapping and does not define the basic simulation length unit. Denoting the experimentally measured translational diffusion coefficient by $D_t$ and the corresponding diffusion coefficient measured in simulation units by $D_t^{\mathrm{sim}}$, we introduce a characteristic time scale $\tau_0$ through
\begin{align}\label{eq:Dtsim}
D_t^{\mathrm{sim}} &= \frac{\tau_0 D_t}{L_0^2}
\end{align}
Solving for $\tau_0$ gives 
\begin{align}\label{eq:tau0}
\tau_0 &= \frac{L_0^2 D_t^{\mathrm{sim}}}{D_t}
\end{align}
Thus physical time is obtained from simulation time by the relation:
\begin{align}\label{eq:tsim_phys}
t_{\mathrm{phys}} &= \tau_0 \ t_{\mathrm{sim}}
\end{align}

\section{Simulation of Hexagonal Colloids}\label{sec:hex_colloids}
Here we describe the simulations done with hexagonal colloids and how we estimated the energy corresponding to the flat sides, from the experiments using total mass-fractions of aggregates $\mathrm{p_{tot}}^{\mathrm{exp}}$.
\subsection{Hexagonal Colloids with two consecutive attractive sides}

\begin{figure*}[h]
    \centering
    \includegraphics[scale=0.925]{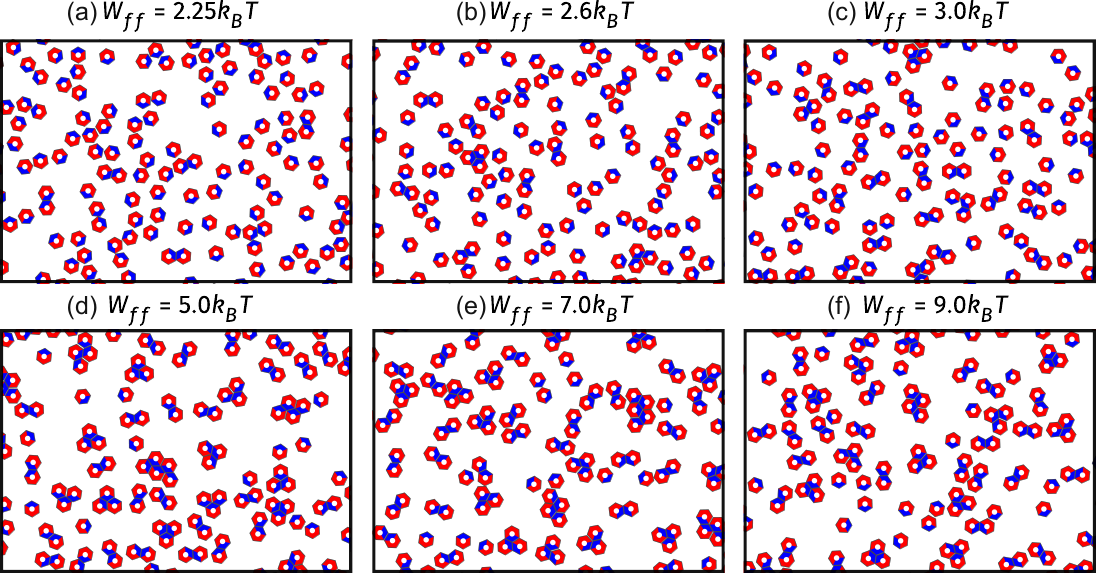}
    \caption{(a)-(f) Snapshots of hexagonal colloid assembly from simulations with consecutive flat sides at the end of 4 hours of experimental time for different values of flat-flat energies. The blue region in the hexagon indicates the attractive/ flat sides and the corrugated/  sides are indicated by the red regions.}
   \label{fig:Fig.S8}
\end{figure*}

We study the formation of self-limiting trimers (T1) as described in \sref{sec:Sec2_6} by using the simulation framework in \sref{sec:model and sim}. Since the hexagonal colloid design in experiments have two consecutive flat sides and the remaining sides which are corrugated; in the simulation, the hexagonal-rigid body is composed of two types of constituent particles. Constituent particles placed along the two consecutive flat sides are assigned the flat type, while constituent particles placed along the remaining four corrugated sides are assigned the corrugated type. Interactions are then defined at the constituent-particle level. Flat constituent particles interact attractively with one another, with target flat-flat interaction strength $W_{ff}$, and as mentioned before in \sref{sec:poly_model}, any interaction involving corrugated constituent particles is taken to be purely repulsive, so that corrugated-corrugated and flat-corrugated contacts do not contribute an attractive binding energy.
Unless stated otherwise, we therefore set only the flat-flat attraction as a variable control parameter and keep all interactions involving corrugated constituent particles non-attractive. Further for comparing the simulation time with experimental time scales, we calculated characteristic particle length $L_0=4.03 \mu m$ for a hexagon of side length $2.5\mu m$ and from \neqref{eq:tau0} characteristic time scale for the hexagonal colloid is calculated to be 54.1 seconds. \nref{fig:Fig.S8} (a-f) shows the resulting assemblies after a duration of $4\text{ hours}$ of experimental time for different values of $W_{ff}$.

\subsection{Estimating the Flat-Flat Energy for the Experiments from Simulations}\label{sec:estimate_Wss}
To infer the effective flat–flat interaction strength $W_{ff}^*$, we first match the experimentally measured total aggregate mass-fraction, $\mathrm{p_{tot}}^{\mathrm{exp}}$ at the corresponding observation time. Twenty simulations with different initial values of velocity is performed for a given value of flat-flat interaction strength $W_{ff}$, and the sample-averaged total aggregate mass fraction $\mathrm{p_{tot}}$ is calculated for each $W_{ff}$. From this, a continuous interpolant using piecewise cubic Hermite interpolation is created. We can then infer $W_{ff}^*$ as the value satisfying 
\begin{align}\label{eq:condition}
\mathrm{p_{tot}}(W_{ff}^*) &= \mathrm{p_{tot}}^{\mathrm{exp}}
\end{align}

\begin{figure*}[h]
    \centering
    \includegraphics[scale=0.925]{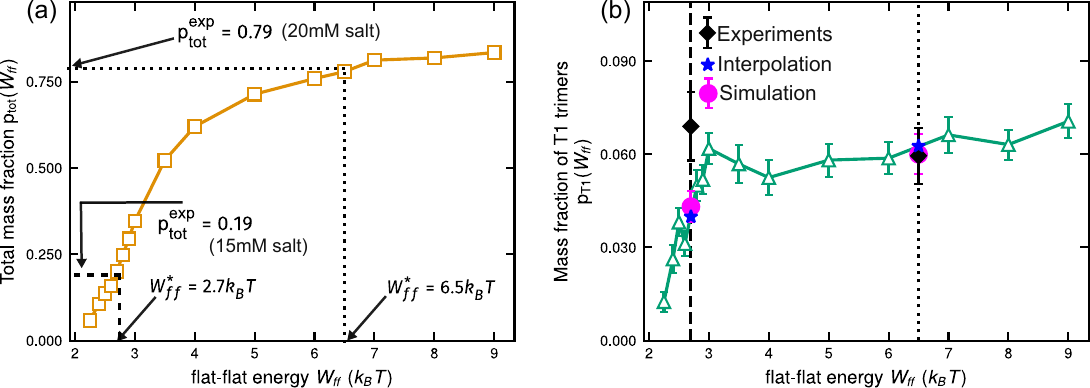}
    \caption{ Mass fractions of aggregates at 4 hours of mapped experimental time from simulation. (a)Total mass fraction of the aggregates as a function of flat-flat energy $W_{ff}$. The horizontal dashed lines indicates the total mass fraction from the hexagonal colloid experiments with different salt concentrations. The vertical dashed lines mark the corresponding values of $W_{ff}$ inferred by interpolation using \neqref{eq:condition}. (b) Mass fraction of self-limiting trimers T1 as a function of flat-flat energy. The blue star and pink circle indicate the corresponding T1 mass fractions obtained at the two inferred interaction strengths, where the star denotes the interpolated value and the circle denotes the directly simulated value. The mass-fraction of T1 trimers obtained from experiments for the two salt concentrations are denoted by the black diamond marker.}
   \label{fig:Fig.S9}
\end{figure*}

When more than one solution exists, the smaller-$W_{ff}$ is chosen, which corresponds to the onset interaction strength required to reproduce the observed extent of aggregation. 
However, for the two cases of salt described in \sref{sec:Sec2_6}, 
we found unique roots --- $6.5k_BT$ and $2.7k_BT$ for high and low salt concentrations respectively, corresponding to experimental total mass fraction of 0.79 and 0.19 (see \nref{fig:Fig.S9}(a)). Having thus fixed $W_{ff}^*$ from the overall aggregate yield alone, we then evaluate the corresponding self-limiting trimer mass fraction $\mathrm{p_\text{T1}}(W_{ff}^*)$ from an independently constructed interpolation of the sample-averaged $\mathrm{p_\text{T1}}(W_{ff})$ data . In this way, $\mathrm{p_{tot}}$ is used to calibrate the overall interaction strength, while $\mathrm{p_\text{T1}}$
provides an independent validation of whether the model also reproduces the experimentally observed structural selectivity. The sample averaged data for $\mathrm{p_\text{T1}}$ is shown in \nref{fig:Fig.S9}. The simulated and interpolated T1 trimer mass fraction denoted by filled circle and star shows very good agreement for the two predicted energies $W_{ff}^* =2.7k_BT$ and $6.5k_BT$. The plot also includes the experimental T1 trimer mass fraction, shown by a diamond marker, and this is in agreement, within error bars, with both the simulated and interpolated values, thus validating this method.


\section{Equilibrium theory of hexagon aggregation}\label{sec:equilibriumtheory} 

We consider the two-dimensional aggregation of hexagonal subunits with two consecutive sticky sides, and aim to compute the mass fraction of aggregates comprising $n$ subunits at equilibrium. This fraction is computed from the concentration $\rho_n$ of aggregates of size $n$, an ideal gas of clusters theory predicts to be~\cite{israelachvili1976theory}
\begin{equation}\label{eq:ideal_gas_of_clusters}
    \rho_n=\nu_nc_1^ne^{-\beta E_n},
\end{equation}
where $\nu_n$ is the number of possible aggregates of size $n$, $c_1$ is the concentration of monomeric subunits at equilibrium and $E_n$ is the binding (free) energy of an aggregate of size $n$.

In \sref{sec:ideal} we rederive the general \neqref{eq:ideal_gas_of_clusters} for pedagogical purposes. In \sref{sec:enumeration} we focus on our specific problem and enumerate all the possible aggregates that our subunits can form to find the expressions of the quantities $\nu_n$ and $E_n$ for our problem. Finally in \sref{sec:fractions} we derive expressions for the equilibrium fractions of each aggregate type.

\subsection{Ideal gas of clusters theory}\label{sec:ideal}

The ideal gas of clusters theory postulates a set of $N$ identical subunits in a volume $V$ (here we consider a two-dimensional volume and thus $V$ has units of surface area) which can aggregate into clusters of different sizes. The system is assumed to be dilute enough for the clusters to be non-interacting. We denote the set of all possible clusters by $\mathcal{C}$. We also denote the size of a cluster $c\in\mathcal{C}$ by $n(c)$, its interaction (free) energy by $E(c)$ and define the particle-wise interaction (free) energy by $e(c)=E(c)/n(c)$. Finally we write the number of clusters of type $c$ in the system as $N_c=V\rho_c$. The grand canonical partition function of the system then reads

\begin{equation}
    \Omega(\beta,\mu) = \sum_{\lbrace N(c)\rbrace}\left\lbrace\prod_{c\in\mathcal{C}}\frac{1}{N_c!}\left[\omega_c(\beta,\mu)\right]^{N_c}\right\rbrace
\end{equation}
where $\beta$ is the inverse temperature, $\mu$ the subunit chemical potential and where the sum runs over all possible partitionings of the subunits into clusters of all types. The partition function for a single cluster of type $c$ reads

\begin{equation}
    \omega_c(\beta,\mu)=V e^{n(c)\beta[\mu-e(c)]}.
\end{equation}

In the thermodynamic limit, these expressions imply a free energy per unit volume

\begin{equation}
    \frac{F(\lbrace\rho_c\rbrace)}{V}=\sum_{c\in\mathcal{C}}\rho_c \left\lbrace n(c)[e(c)-\mu] +[\ln\rho_c-1]\right\rbrace,
\end{equation}
which is minimized for a set of concentrations given by

\begin{equation}\label{eq:rhoc}
    \rho_c=e^{n(c)\beta[\mu-e(c)]}=\left[c_1 e^{-\beta e(c)}\right]^{n(c)},
\end{equation}
where we have set $e(1)=0$ by convention. In a system where the total number of subunits is fixed, $\mu$ (or equivalently $c_1=\exp(\beta\mu)$) is an unknown quantity that can be computed by solving the equation

\begin{equation}\label{eq:concentration}
    N/V = \sum_{c\in\mathcal{C}}n(c)\rho_c = \sum_{c\in\mathcal{C}}n(c)\left[c_1 e^{-\beta e(c)}\right]^{n(c)},
\end{equation}
which expresses the constraint that the sum of the masses of all cluster types equal the total mass.

To derive \neqref{eq:ideal_gas_of_clusters} from these results, we denote by $\mathcal{C}_n$ the set of clusters comprising $n$ subunits. In the case where all clusters in $\mathcal{C}_n$ have the same energy, we can use \neqref{eq:rhoc} to write $\rho_n=\sum_{c\in\mathcal{C}_n}\rho_c$ under the form given in \neqref{eq:ideal_gas_of_clusters}, where $\nu_n$ denotes the number of clusters in $\mathcal{C}_n$.

\subsection{Enumeration of possible aggregates}\label{sec:enumeration}

\begin{figure}[t]
    \centering
    \includegraphics[scale=0.925]{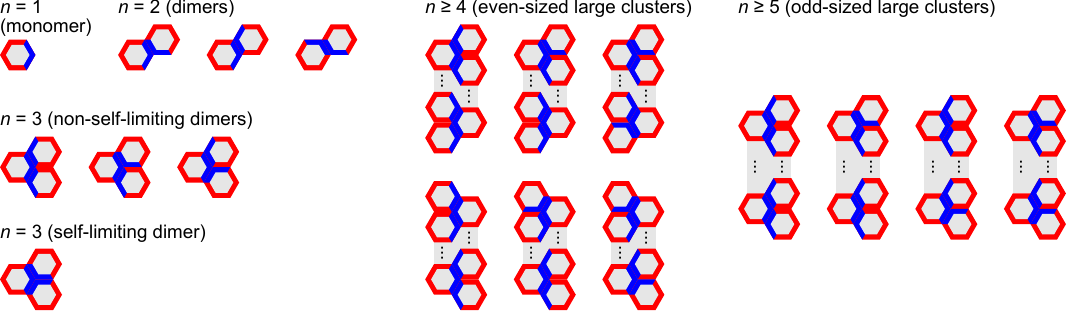}
    \caption{
    Illustration of all possible cluster configurations for our subunits.
    \emph{Monomer}: We consider hexagons with two consecutive sticky (\emph{blue}) and four non-sticky (\emph{red}) sides.
    \emph{Dimers}: Dimeric clusters differ not by their overall shape, but by the orientation of their subunits. While the first cluster shown here is achiral (the contact line between the two subunits forms an axis of symmetry), the last two are chiral and are mirror images of each other. Any possible dimer can be obtained by rotating one of these three configurations.
    \emph{Self-limiting trimer}: This is the only cluster where all sticky sides are in contact with another sticky side, and therefore the ground state of the system.
    \emph{Non-self-limiting trimers}: These configurations are obtained from the previous one by rotating one or two subunits away from one of their neighbors. Every subunit must retain at least one bond to the rest of the cluster, lest the cluster breaks into two smaller ones.
    \emph{Even-sized large clusters}: All unbroken large clusters must be fiber-like due to our subunit design. The differences between clusters of the same size thus only reside in the structure of their ends. Unlike odd clusters, even-sized large clusters have an overall chiral shape even when the orientations of their subunits are not taken into account. The first row represents one possible chirality. It includes the case where two, one or zero sticky sides are facing away from the cluster and available for further binding. The clusters of the bottom row are the mirror images of these configurations.
    \emph{Odd-sized large clusters}: Unlike in the previous case, these clusters have an achiral overall shape. This implies that there is only one configuration with two available sticky sides, and one with zero. However there still are two configurations with one available side. Indeed the configuration with the bottom subunit unavailable cannot be obtained by an overall rotation of the configuration with the top subunit unavailable.
    }
    \label{fig:idc}
\end{figure}

We now apply the general considerations detailed above to the specific case of our hexagons with two consecutive sticky sides. We denote by $\epsilon$ the (free) energy associated with the binding of two such sides, with the convention that $\epsilon<0$ if the binding is indeed favorable. Contacts between subunits involving either one or two non-sticky sides are neither favored nor penalized and do not contribute to $E_c$. As shown in \nref{fig:idc}, all clusters but the self-limiting trimer have one fewer bonds than they have subunits, implying

\begin{equation}
    E_n = (n-1)\epsilon, \quad\text{except for self-limiting trimers (denoted by T1 in the main text), where}\quad E_\text{T1}=3\epsilon.
\end{equation}
The same figure shows all possible cluster morphologies associated with each value of $n$, implying

\begin{itemize}
    \item $\nu_1=1$: trivially there is only one way to make a monomer
    \item $\nu_2=3$
    \item $\nu_3=3$: there are three non-self-limiting trimers in addition to T1.
    \item $\nu_{2k}=4$ for clusters with an even number $n=2k\geq 4$ of subunits.
    \item $\nu_{2k+1}=6$ for clusters with an odd number $n=2k+1\geq 5$ of subunits.
\end{itemize}

This provides us with all values of $E_c$ and $\nu_n$, as required to build the full equilibrium gas of clusters theory.

\subsection{Fraction of each aggregate type}\label{sec:fractions}

Inserting these results into \neqref{eq:rhoc}, \eqref{eq:concentration} and assuming a fixed subunit concentration $N/V$, we obtain the following equation for the unknown equilibrium concentration of monomers $c_1$:

\begin{align}\label{eq:c1_equation}
    \frac{N}{V} &= c_1 + 3 \left(2c_1^2e^{-\beta\epsilon}\right) + 3c_1^3e^{-3\beta\epsilon} + 3 \left(3c_1^3e^{-2\beta\epsilon}\right)
    + \sum_{n\geq 4\text{ even}}6nc_1^ne^{-(n-1)\beta\epsilon}+ \sum_{n\geq 5\text{ odd}}4nc_1^ne^{-(n-1)\beta\epsilon}\nonumber\\
    &= -3c_1\left(1+c_1e^{-\beta\epsilon}\right)^2+3c_1^3e^{-3\beta\epsilon}+\frac{4c_1}{\left(1-c_1e^{-\beta\epsilon}\right)^2}+\frac{4c_1^2e^{-\beta\epsilon}}{1-\left(c_1e^{-\beta\epsilon}\right)^2},
\end{align}
where the successive terms of the right-hand-side of the first equality respectively correspond to the categories highlighted in \nref{fig:idc}.

To compute the fraction $\phi_\text{T1}$ of all subunits that belong to a self-limited trimer at equilibrium, we numerically solve \neqref{eq:c1_equation} for $c_1$ using the values of $N/V$ and $\beta\epsilon$ corresponding to the experimental or simulated system under consideration. We then use the result to compute

\begin{equation}
    \phi_\text{T1}=\frac{\rho_\text{T1}}{N/V}=\frac{3\left(c_1e^{-\beta\epsilon}\right)^3}{N/V}.
\end{equation}
Likewise, the fraction of subunits comprised in non-T1 aggregates of size $n$ reads $\phi_n=\nu_nc_1^ne^{-(n-1)\beta\epsilon}/(N/V)$. We use these expressions to derive the theoretical results presented in \nref{fig:Fig.6} of the main text.

\section{Supplementary movies}

\noindent \textbf{\href{https://drive.google.com/file/d/1xgrGh7-xrIOIv-DaFNQMR6bR3fQn2a02/view?usp=drive_link}{Movie-S1}} : Selective self-assembly of 3D printed rectangular colloids on their shorter sides through depletion interaction, resulting 1D chains. The movie was recorded at 2 frames per minute and is played at 10 frames per second. Duration of the movie in real time is 2 hours. 

\noindent \textbf{\href{https://drive.google.com/file/d/16h4zoL5j9vGWqGKzIEgH8KL7lekt0EsO/view?usp=drive_link}{Movie-S2}} : Stabilization of self-limited trimers (T1) and destabilization of other aggregates. The movie was recorded at 2 frames per minute and is played at 10 frames per second. Duration of the movie in real time is 50 minutes.

\bibliographystyle{unsrt}
\bibliography{Bib}

\end{document}